\documentclass{SciPost}

\hypersetup{
    colorlinks,
    linkcolor={red!50!black},
    citecolor={blue!50!black},
    urlcolor={blue!80!black}
}

\usepackage[bitstream-charter]{mathdesign}
\usepackage{slashed}
\usepackage{soul}
\usepackage{orcidlink}
\newcommand*{\La}{{\cal{L}}}
\newcommand*{\no}{\noindent}
\newcommand*{\bea}{\begin{eqnarray}}
\newcommand*{\eea}{\end{eqnarray}}
\newcommand*{\be}{\begin{equation}}
\newcommand*{\ee}{\end{equation}}

\newcommand*{\pref}[1]{(\ref{#1})}

\newcommand*{\mn}{{\mu\nu}}

\newcommand*{\prefr}[2]{(\ref{#1}-\ref{#2})} 
\newcommand*{\nn}{\nonumber}

\newcommand*{\tr}{\mathrm{tr}}

\newcommand{\bma}{\begin{pmatrix}}
\newcommand{\ema}{\end{pmatrix}}

\newcommand*{\op}{{{\cal O}}}
\newcommand*{\la}{\left\langle}
\newcommand*{\ra}{\right\rangle}

\DeclareSymbolFont{usualmathcal}{OMS}{cmsy}{m}{n}
\DeclareSymbolFontAlphabet{\mathcal}{usualmathcal}

\fancypagestyle{SPstyle}{
\fancyhf{}
\lhead{\colorbox{scipostblue}{\bf \color{white} ~SciPost Physics Core }}
\rhead{{\bf \color{scipostdeepblue} ~Submission }}

\fancyfoot[C]{\textbf{\thepage}}
}

\begin{document}

\pagestyle{SPstyle}

\begin{center}{\Large \textbf{\color{scipostdeepblue}{Weak (quasi-)PDFs from the lattice}}}\end{center}


\begin{center}\textbf{
Patrick Jenny\orcidlink{0000-0003-1704-5224},
Axel Maas\orcidlink{0000-0002-4621-2151}\textsuperscript{$\star$},
Sofie Martins\orcidlink{0000-0001-5428-8925}\textsuperscript{$\oplus$}, and
Georg Wieland\orcidlink{0009-0003-1777-4563}\textsuperscript{$\dagger$}
}\end{center}

\begin{center}
Institute of Physics, University of Graz, NAWI Graz, Universit\"atsplatz 5, 8010 Graz
\\[\baselineskip]
$\star$ \href{mailto:axel.maas@uni-graz.at}{\small axel.maas@uni-graz.at}\,,\quad
$\oplus$ \href{mailto:sofie.martins@uni-graz.at}{\small sofie.martins@uni-graz.at}\,,\quad
$\dagger$ \href{mailto:georg.wieland@uni-graz.at}{\small georg.wieland@uni-graz.at}
\end{center}

\section*{\color{scipostdeepblue}{Abstract}}
\textbf{\boldmath{
Future colliders will be able to probe the electroweak substructure of particles. An accurate description of this substructure may necessitate electroweak PDFs, similarly to QCD. We investigate therefore the Higgs and $W$/$Z$ PDFs in a reduced Standard Model setup using lattice methods, determining the quasi-PDFs, and, to lowest order, the resulting PDFs. While uncertainties remain, the results show interesting structures in line with expectations from gauge invariance.\\
}}

\no Preprint number: COMETA-2026-46

\vspace{\baselineskip}

\noindent\textcolor{white!90!black}{%
\fbox{\parbox{0.975\linewidth}{%
\textcolor{white!40!black}{\begin{tabular}{lr}%
  \begin{minipage}{0.6\textwidth}%
    {\small Copyright attribution to authors. \newline
    This work is a submission to SciPost Physics Core. \newline
    License information to appear upon publication. \newline
    Publication information to appear upon publication.}
  \end{minipage} & \begin{minipage}{0.4\textwidth}
    {\small Received Date \newline Accepted Date \newline Published Date}%
  \end{minipage}
\end{tabular}}
}}
}



\vspace{10pt}
\noindent\rule{\textwidth}{1pt}
\tableofcontents
\noindent\rule{\textwidth}{1pt}
\vspace{10pt}

\section{Introduction}\label{s:intro}

Using parton distribution functions (PDFs) to describe the structure of bound states in collision processes of hadrons has a long and successful tradition~\cite{Dissertori:2003pj,BeiglboCk:2006lfa}. As higher energies become experimentally accessible at future colliders, such as the currently favored next large-scale collider \textit{FCC}~\cite{FCC:2025lpp,deBlas:2025gyz}, on-shell sea Higgs and weak gauge bosons become relevant, contributing substantially to cross sections~\cite{Bauer:2017isx,deBlas:2025gyz,MoralFigueroa:2025cev}. Correspondingly, we need to understand contributions to PDFs originating from the electroweak sector~\cite{Bauer:2017isx}. These contributions are thought to be accessible through perturbative calculations, similar to QED PDFs~\cite{BeiglboCk:2006lfa}, as the only contributions come from the sea, and splitting functions are suppressed by the small weak coupling. However, this holds only under the assumption that the structure of the asymptotic states is fully understood.

In the perturbative construction the electroweak asymptotic states do not need to be gauge invariant with respect to the weak interaction, since in the presence of the Brout-Englert-Higgs (BEH) effect, the gauge symmetry is broken by a gauge-variant vacuum. That the perturbatively constructed gauge-variant asymptotic states are the elementary ones is then responsible for the lack of valence contributions to the aforementioned PDFs. But this is not the complete picture.

In fact, the gauge symmetry is actually not literally broken, a consequene of Elitzur's theorem~\cite{Elitzur:1975im}, but merely hidden. The precise agreement of experimental measurements and perturbative phenomenological calculations is then not trivial, but can be explained by the Fr{\"o}hlich-Morchio-Strocchi (FMS) mechanism ~\cite{Frohlich:1980gj,Frohlich:1981yi}, and may not hold always. In fact, the FMS mechanism has been used to predict deviations~\cite{Maas:2017wzi,Dudal:2020uwb,Maas:2020kda} at sufficient precision. Hence, physical asymptotic states need to be weakly gauge-invariant. For particles like the proton or the electron, this implies the existence of an electroweak valence contribution of the Higgs~\cite{Frohlich:1980gj,Frohlich:1981yi,Egger:2017tkd}, which is consistent with current data~\cite{Fernbach:2020tpa}.

Future colliders may be sensitive to this effect. As manifestly invariant asymptotic states also yield a different composition of the PDFs~\cite{Maas:2022gdb} than in perturbation theory, it is worthwhile to take this possibility into account explicitly. Especially as data-driven extractions of the PDFs can hide new physics~\cite{Gao:2017yyd}, and thus potentially also such effects. In fact, it may be that the full electroweak PDF structure can still be calculated by a suitable augmentation of perturbation theory~\cite{Maas:2017wzi,Maas:2020kda,Dudal:2020uwb,Maas:2024hcn}. However, to avoid an unnecessary assumption at this stage, here a different route will be taken.

It has recently become possible to determine PDFs using first-principles lattice methods~\cite{Ji:2013dva,Cichy:2018mum}. The approach applies analogously to electroweak physics. However, a full Standard Model computation is hampered by the fact that parity violation in gauge theories is obstructed by the lattice regularization~\cite{Hasenfratz:2007dp}, and the large scale separations are computationally unfeasible. Thus, to test PDFs, we turn here to the use of a truncated Standard Model, which contains only the Higgs and the W/Z gauge bosons. This theory has been studied on the lattice in great detail~\cite{Maas:2017wzi}, and the predicted deviations due to the compositeness of the asymptotic states from the perturbative results are readily available~\cite{Maas:2018ska,Jenny:2022atm,Martins:2026eni,Maas:2020kda,Dudal:2020uwb,Maas:2024hcn}. We thus use this simplified theory here to determine the PDFs using lattice methods, providing insight into the structure of the composite gauge-invariant asymptotic states. Our results are compatible with the ones favored by general arguments~\cite{Egger:2017tkd} and constraints from data~\cite{Fernbach:2020tpa}. Future work may be concerned with computing PDFs from augmented perturbation theory (APT), incorporating the gauge invariance, and comparing them with the non-perturbative results in this work. This will ultimately pave the way to use APT to calculate the electroweak PDFs for the full standard model, rather than the present proxy.

We will also supplement the precision results with a first investigation which also include unquenched vectorial fermions, an improved proxy for the Standard Model~\cite{Afferrante:2020fhd,Wieland:2025tog}. Some preliminary results can be found in~\cite{Martins:2026eni}.

The theoretical setup of our calculation will be given in section~\ref{s:theory}, in which we also discuss the details of the PDFs. Lattice simulation details can be found in section~\ref{s:lattice}. While the lattice techniques are standard, the application to weak physics is new. The results for the PDFs, including the analysis of systematic lattice artifacts, are given in section~\ref{s:lpdf}. Our final results are presented in section~\ref{s:pdf} with a summary and conclusion in section~\ref{s:conclusion}. 

\section{Theoretical setup}\label{s:theory}

\subsection{Theory}

The theory we study comprises an $\mathrm{SU(2)_w}$ Yang-Mills theory with gauge bosons $W$ coupled to a scalar field $X$ in the fundamental representation and two generations of degenerate vectorial fermions $\psi$ in the fundamental representation. The Lagrangian of this theory is given by 
\be
\La = -\dfrac{1}{4} W^{a}_{\mu \nu} W^{a \,\mu \nu} + \dfrac{1}{2} \tr \left[ (D_\mu X)^\dag (D^\mu X) \right] - \frac{\lambda}{4} \left( \tr (X^\dag X)-v^2  \right)^{2} + \bar{\psi}\,\big(\mathrm{i}\slashed{D}-m_\psi\big)\,\psi\,.\label{lag}
\ee
It is thus structurally similar to the Standard Model lepton sector, except that the vectorial nature of the fermions allow tree-level masses. This theory has several symmetries besides the $\mathrm{SU(2)_w}$ gauge symmetry: A global $\mathrm{SU(2)_c}$ symmetry of the scalar field $X$ and an $\mathrm{SU(2)_g}$ generational symmetry $\psi$ of the fermions for degenerate mass. As the gauge symmetry $\mathrm{SU(2)_w}$ yields a pseudoreal fundamental representation, the generation symmetry combines with the chiral symmetry to an enlarged $\mathrm{SU(4)_{\psi}}$, known as Pauli-G\"ursey symmetry~\cite{Kogut:2000ek}. If $m_\psi\neq 0$, this symmetry is explicitly broken to an $\mathrm{Sp(4)_\psi}$ symmetry. Besides this explicit breaking, the global symmetries remain intact, and can thus be used to classify states. Below, most of the time, the theory~\pref{lag} will be used without the fermions.

As noted in the introduction, the physical spectrum of the theory needs to be manifestly invariant under the gauge symmetry. Thus, states can only be classified according to their global symmetries, as well as $J^P$. A separate charge symmetry is, due to the pseudoreality, absent even for uncharged states. A detailed discussion of this can be found in~\cite{Afferrante:2020fhd}.

Here, we consider two quantum number channels, the uncharged scalar $0^+$ and the custodial triplet $1^-$. These correspond to the physical Higgs and the physical W and Z bosons~\cite{Frohlich:1980gj,Frohlich:1981yi,Maas:2017wzi}. The simplest composite operators for these quantum numbers are
\bea
O_0&=&\det X\label{op1}\,,\\
O_1^{a\mu}&=&\tr\left(\tau^a X^\dagger D^\mu X\right)\label{op2}\,,
\eea
\no where $\tau$ are the Pauli matrices acting in $\mathrm{SU(2)_c}$. We note that the vector operator will generally have three polarization directions. We will consider here the cases, where the operator $\op_1$ is projected longitudinally and transversely with respect to the momentum separately, indicating the two possibilities as $\parallel$ and $\perp$ respectively. With our momentum introduced later as $(P_0,0,0,P_3)$, this implies that we either average over the $\mu=1,2$  polarizations or just consider the $\mu=3$ case. Since the vector operator is gauge invariant, all polarization directions are physical on-shell. Moreover, in our reduced Standard Model setups, there are stable states in both channels that produce on-shell asymptotic states, whose spin and spin components are well defined observable quantities. A brief exploratory discussion on how this relates to the gauge-dependent polarizations of the gauge bosons can be found in~\cite{Maas:2018ska}.

\subsection{PDFs}

In general, any PDFs will be derived from matrix elements of the structure~\cite{Cichy:2018mum,Ji:2013dva}
\be
C_3(P,z)=\la O_i^\dagger(P) J(z,0) O_i(P)\ra\,,\label{c3}
\ee
\no where the $O_i$ are any of~\prefr{op1}{op2}, and represent the interacting state of the corresponding quantum number and momentum $P=(P_0,0,0,P_3)$. The quantity $J$ is a gauge-invariant bi-local current operator, potentially carrying some charge, connecting the points $0$ and $z$ along some path. A possibility are two terminating operators at the location of $O_i$ and $O_i^\dagger$ and a connecting Wilson line.

As a PDF is build from the current $J$ of a conserved charge, it can be interpreted as the momentum distribution of that charge in the given particle described by $O_i$. In case the charge is carried by a specific species of particle, this becomes equivalent to the momentum distribution of this type of particle inside the parent particle. This interpretation is especially applying to the usual context of QCD, where all elementary particles are distinguished by global quantum numbers. These map one-to-one on physical currents, the six quark flavor currents and an unflavored spin-1 current for the gluon. The corresponding currents are then dominated at lowest order by the elementary particles, allowing this one-to-one interpretation. Of course, a literal interpretation as a probability is not possible beyond tree-level~\cite{BeiglboCk:2006lfa}. 

In our context, the situation is drastically different. The physical quantum numbers of~\prefr{op1}{op2} are not the same as those of the elementary ones. This makes a similar interpretation difficult, and needs some explanation. In particular, this will substantially affects the choice of currents.

In its very definition, the expression~\pref{c3} is build from gauge-invariant operators. In QCD, the connection to the elementary gauge-dependent constituents is then usually drawn in the following way: The inserted bilocal current $J$ has for quarks, e.\ g., the form~\cite{Cichy:2018mum}
\be
J=\bar{q}(x) W({\cal C}_{x\to y}) q(y)\,,\label{melement}
\ee
\no where $W$ is a Wilson line along a path ${\cal C}_{x\to y}$ from $x$ to $y$, and the terminating quark fields $q$ are here taken to be of the same flavor. At leading order, $W\sim\bf{1}$, and thus~\pref{melement} provides the correlation of a quark inside each of the asymptotic states with fixed flavor. Likewise, the usual field-strength tensor will yield at leading order a gluon. As~\pref{melement} is gauge-invariant, it averages over all possible colors of the quarks.

In the present case, the physical states for the $W$ and $Z$ are described by a custodial triplet~\pref{op2}. However, the elementary ones are a gauge triplet and carry no custodial charge. Thus, building the current on~\pref{op2} will not yield an identification with the $W$ and $Z$, but rather with multiple Higgs to create the adjoint charge. This is not what is needed for a perturbative approach\footnote{While there is a one-to-one mapping for the spectrum between gauge states and custodial states for the vector~\cite{Frohlich:1980gj,Maas:2017wzi}, this is not true for the currents in~\pref{melement}. Consequently, constructing a custodial vector current gave us a zero result within errors. This is like looking for a hadron in another hadron by means of a PDF. While such a thing is conceptually possible, it is not expected to be a significant contribution, and thus a negligible signal is expected.}. Thus, to adequately describe the $W$ and $Z$ in a PDF language will require to treat them in the same way as gluons in QCD, i.\ e.\ by a $J$ built from the field strength tensors. Note that in the present case $W$ and $Z$ are the same, because we do not include hypercharge. Thus, summing over $W$ and $Z$ equally in the perturbative treatment is for now adequate.

This leaves the Higgs, which splits perturbatively into the fluctuation mode and the pseudo-Goldstone bosons contained in the Higgs field. By the Goldstone-boson equivalence theorem, the latter are expected to be related to the longitudinal polarization of the vector bosons. Doing the calculation in a suitable gauge, e.\ g.\ unitary gauge, all of their effects would be contained in the cases listed so far. If the aim is to use the PDFs in a calculation where the pseudo-Goldstone bosons remain dynamical, it may be necessary to also build from the Higgs fields a current in the adjoint of the custodial group. However, this is beyond the current scope. Here, we will concentrate on the scalar singlet current, to describe the fluctuation Higgs field, making contact to previous phenomenological investigations~\cite{Egger:2017tkd,Fernbach:2020tpa}.

In practice, the operator $J$ is not unique, as usual. However, it has been argued that the results are universal~\cite{Zhang:2018diq}. In practice, we build our operators from the following basic operators
\bea
F_{0^+}(\tau,z_3)&=&\tr \left( X^\dagger(z)W(z,z_0)X(z_0) \right)\,,\label{f0}\\
F_{1^-}^1&=&w_{zx}+w_{zy}\,,\label{f1}\\
F_{1^-}^2&=&w_{zx}+w_{zy}+w_{zt}\,,\label{f2}\\
F_{1^-}^3&=&w_{tx}+w_{ty}-2w_{xy}=F_{1^-}^4-2F_{1^-}^5\,,\\
F_{1^-}^{4}&=&w_{tx}+w_{ty}\,,\label{f4}\\
F_{1^-}^{5}&=&w_{xy}\,,\label{f5}\\
w_\mn(\tau,z_3)&=&\tr \left(W_\mn(z)W(z,z_0)W_\mn(z_0)W(0,z_0) \right)\,,\nn\\
z&=&(\tau,0,0,z_3)\,,\nn\\
z_0&=&(\tau,0,0,0)\,,\nn
\eea
\no where $W$ is a straight Wilson line connecting the two points $z$ and $z_0$, which are at the same time $\tau$. As we choose the straight line along the third axis the only explicit dependency is on the difference in this component, called $z_3$. We note that the operators $F_{1^-}^i$ are the same as the ones used for the gluon PDFs in~\cite{Fan:2020cpa,HadStruc:2021wmh,Balitsky:2019krf,Good:2024iur}, as discussed above. Only $F_{0^+}$ is genuine to the present context. As we found that in our case in most setups $F_{1^-}^4$ and $F_{1^-}^5$ cancel within statistical errors in $F_{1^-}^3$, we investigated each of them separately. We found that they yield similar results to $F_{1^-}^1$ and $F_{1^-}^2$ individually.

Together with the possible polarizations, we studied in total 18(=3 external states$\times$6 currents) cases. We studied these in three distinct setups, two are pure gauge-scalar systems without fermions, which differ (slightly) in terms of the parameters. The third set of ensembles includes dynamical fermions, but with fewer and smaller volumes and reduced statistics due to larger computational cost. Note that probing the fermion as a state is more involved due to the symmetry structure~\cite{Maas:2022gdb}. This will be addressed in the future. In total, we determine 54 different (quasi-)PDFs.

The PDFs are obtained from the matrix element~\pref{c3} through the intermediate step of quasi-PDFs. To this end, the normalized operator
\bea
R^{OF}(t,\tau,P_3,z_3)&=&\frac{C_3^{OF}(t,\tau,P_3,z_3)}{C_2^O(P_3,t)}\,,\nn\\
C_3^{OF}&=&\la O^\dagger(P,t) F(z,0,\tau) O(P,0)\ra\,,\nn\\
C_2^{O}&=&\la O^\dagger(t,P_3) O(0,P_3)\ra\nn
\eea
\no is build from any choice of~\prefr{op1}{op2} for $O$, separately for each polarization, and~\prefr{f0}{f5} for $F$. The momentum dependence for the operators $O$ is introduced by performing a Fourier transformation with respect to the spatial directions. The time difference, measured by $t$, is used to pull apart the probed particles, such that initial state and final state are far away from the probe time $\tau$, and should therefore be dominated by the asymptotic states. Ideally, $0\ll\tau\ll t$.

Assuming $0\ll\tau\ll t$, this behaves as~\cite{Cichy:2018mum}
\be
\lim_{t\to \infty}R^{OF}(t,\tau,P_3,z)=K^{O}(P_3)M^{OF}(P_3,z)\,,\nn
\ee
\no where $K$ is a known function, a kinematic normalization. For~\pref{f0} it is $1/(2E)$, where $E$ is the energy associated with the probed particle, and for~\prefr{f1}{f5} it is the inverse gauge coupling $1/g$. Both will be absorbed in the following, as we will eventually normalize the resulting PDFs.

The function $M$ is thus the one of interest here, and can be obtained as
\be
\frac{d}{dt}\int_{\epsilon}^{t-\epsilon}d\tau \, R^{OF}(t,\tau,P_3,z)=M^{OF}(P_3,z)+{\cal O}(t)\,.\label{mfunc}
\ee
\no In turn, the quasi-PDF $Q$ is then defined as~\cite{Cichy:2018mum}
\be
Q^{OF}(x,P_3)=N^{F}(xP_3)\int\frac{dz}{2\pi}e^{-izxP_3}M^{OF}(P_3,z)\label{quasipdf}
\ee
\no with $N^{0^+}=1/2$~\cite{Wang:2017qyg} and $N^{1^-}=1/(xP_3)$~\cite{HadStruc:2021wmh} once more normalization factors.

The quasi-PDF relates to the PDF $q$ as~\cite{Cichy:2018mum,Ji:2013dva}
\be
Q^{OF}(x,P_3)=\int_{-1}^1\frac{dy}{|y|}D^{OF}(x,y,P_3,\mu) q^{OF}(y,\mu)\,,\nn
\ee
\no where $D^{OF}$ is a perturbative expression, which thus introduces a factorization scale $\mu$. This function $D^{OF}$ takes the form
\be
D^{OF}x,y,P_3,\mu)=\delta\left(1-\frac{x}{y}\right)+\alpha d^{OF}(x,y,P_3,\mu)\,,\nn 
\ee
\no with a perturbatively calculable function $d^{OF}$, which becomes less relevant for $P_3\to\infty$, and $\alpha$ is the relevant coupling constant. As in our present case $\alpha$ is much smaller than in QCD, it can be expected that the correction due to $d^{OF}$ is much smaller as in the QCD case. We thus take here the approximation
\be
q^{OF}(x)=\lim_{P_3\to\infty} Q^{OF}(x,P_3)\,.\label{pdf}
\ee
\no Note that this implies that the factorization scale dependence is lost, and the so determined PDF will be at most useful at tree-level.

In the context of APT calculations, this will likely suffice for some time to come. However, it should be noted that we refer here to an infinite momentum frame for the PDFs, neglecting masses. It is therefore likely that our PDFs would be appropriate for very high-energy collisions. At low energies, of order a few hundred GeVs, the PDFs will likely be much more similar to their tree-level counterparts discussed in~\cite{Egger:2017tkd}, i.\ e.\ trivial
\bea
H_H&=&\delta(1-x)\,,\label{tlpdf1}\\
H_W&=&0\,,\label{tlpdf2}\\
W_W&=&\delta(1-x)\,,\\
W_H&=&0\,,\label{tlpdf4}
\eea
\no irrespective of polarization. This is in line with measurements of cross sections in this theory using lattice simulations at very low energies~\cite{Jenny:2022atm}: Up to additional threshold effects due to extension of the states, they can be described well with the tree-level PDFs~\prefr{tlpdf1}{tlpdf4}. One should therefore expect a $Q$-evolution from~\prefr{tlpdf1}{tlpdf4} to the ones presented here. This is necessarily left for future work.

\subsection{Normalization}\label{s:normalization}

For a final result, it is necessary to normalize the results for $P_3\to\infty$. Assuming that all other PDFs, which are not determined in this work, are negligible, there is always the energy sum rule
\be
\int dx x\left(\sum q^i(x)\right)=1\,.\nn
\ee
\no However, this requires a second sum rule for normalization, which is less simple to identify.

For the scalar, there is no conserved quantum number available, and thus no obvious sum rule. However, we will adopt a constituent counting rule as
\be
\int dx H_H(x)=1\,.\label{higgssumrule}
\ee
\no The reasoning here is the FMS mechanism. When working in a suitable fixed gauge, the operator~\pref{op1} can be rewritten as~\cite{Frohlich:1980gj,Frohlich:1981yi,Maas:2017wzi}
\be
\det X =v^2+v\Re\tr\eta+\tr \left(\eta^\dagger\eta\right)\label{higgs}
\ee
\no with $X=v{\bf 1}+\eta$, where $v$ is the Higgs vacuum expectation value and $\eta$ the fluctuation field. Hence, on-shell, the scalar is resembled by a single elementary Higgs~\cite{Dudal:2020uwb,Maas:2020kda,Maas:2024hcn}. As a result, when neglecting terms of order $\eta/v$, the PDFs for the Higgs will be just the tree-level PDFs~\pref{tlpdf1} and~\pref{tlpdf2}~\cite{Egger:2017tkd}, justifying (and satisfying) this sum rule.

For the vector, the situation is different, as there is a conserved quantum number, the custodial charge. Now, for the same reason as in~\pref{higgs}, follows
\be
\tr \left( \tau^a X^\dagger D_\mu X \right)=v^2 W_\mu^a+v\left(\tr \left(\tau^a D_\mu \eta \right) +W_\mu^b\tr \left( \tau^a\eta \tau^b \right) \right)+\tr \left(\eta^\dagger D_\mu \eta \right)\,.\label{vector}
\ee
\no Thus, to leading order, this suggests that the vector can be approximated by a gauge field. This would suggest a similar sum rule as~\pref{higgssumrule},
\be
\int dx W_W(x)=1\,,\label{custsr}
\ee
\no for both transverse and longitudinal polarization separately. However, this is problematic. On the one hand, e.\ g., in Landau gauge the $W$ has no longitudinal degrees of freedom. Thus, the elementary field will not contribute to the longitudinally polarized vector -- at least not at leading order. In addition, in~\pref{vector}, the custodial charge is actually carried by the vacuum condensate $v$, as well as $\eta$. Thus, it is more appropriate to require the sum rule
\be
\int dx W_H(x)=2\label{custer}
\ee
\no for the two units of fundamental charge contributed to the state by the Higgs field. This sum rule would not be satisfied by~\prefr{tlpdf1}{tlpdf4}. However, setting
\be
W_H=\frac{1}{2}\delta(x)\label{tlpdf5}
\ee
\no would do so. Such a PDF does not contribute at leading order to (elastic) cross sections~\cite{Egger:2017tkd}, and is thus still compatible with existing (lattice) data. It will therefore be adopted here. The argument holds doubly true at high energies, as there the Higgs partons are expected to be treated on equal footing, i.\ e.\ all terms in~\pref{vector} should be equally important~\cite{Egger:2017tkd,Maas:2022gdb}. Then the fact that the custodial charge is carried by the Higgs partons needs to be fully evident. However, the crossover between both regions may be fairly involved, and may affect the effective choice of PDFs~\cite{Maas:2022gdb}.

This also avoids a technical problem. The normalization for the vector in~\pref{quasipdf} yields a trivial $1/x$ behavior for the gauge boson quasi-PDFs. For a sum rule like~\pref{custsr} to hold, this would need to be suppressed by the matrix element. As will be seen, our results do not (yet) see a suppression at small $x$. While this may be just a lattice artifact, this would require a regularization to employ a sum rule like~\pref{custsr}.

\section{Lattice setup}\label{s:lattice}

\subsection{Parameter sets}

The configurations for this project have been created using the HiRep code~\cite{DelDebbio:2008zf} with an extension for scalar fields~\cite{Hansen:2017mrt} in the manner discussed in~\cite{Maas:2026dlq}, to which we refer for full details.

\begin{table}
\centering
    \begin{tabular}{|c|c|c|c|c|c|c|c|c|}
    \hline
     Name & $\beta$ & $\kappa$ & $\gamma$ & $\kappa_F$ & $a^{-1}$ [GeV] & $m_h$ [GeV] & $N_g$ & $m_\psi$ [GeV] \cr
     \hline
     H & 2.7984 & 0.2984 & 1.317 & & 287 & 148 & 0 & \cr
     L & 8.0000 & 0.1310 & 0.0033 & & 400 & 125 & 0 & \cr
     F & 8.0000 & 0.1310 & 0.0033 & 0.3103 & 400 & 125 & 2 & 128 \cr 
     \hline
     \end{tabular}
     \vskip10pt
    \begin{tabular}{|c|c|c|c|c|}
    \hline
     Name & $v$ [GeV] & $g$ & $\alpha_W$ & $\lambda$ \cr
     \hline
     H & 68.1 & 1.18 & 0.111 & 15.0 \cr
     L & 149 & 0.54 & 0.023 & 0.69 \cr
     F & 149 & 0.54 & 0.023 & 0.69 \cr 
     SM & 250 & 0.32 & 0.0081 & 0.25 \cr
     \hline
     \end{tabular}
    \caption{\label{tab:parameter_sets}The upper table lists the simulation parameters and sets. The lattice spacing $a$ has been set by requiring that $m_W=80.375$ GeV. The lattice spacing and the masses have been determined in~\cite{Maas:2014pba,Jenny:2022atm} for set $H$, in~\cite{Wurtz:2013ova} for set $L$, and in~\cite{Maas:2026dlq} for set $F$. The lower table lists the tree-level parameters in the renormalization scheme defined in~\cite{Maas:2026dlq}, for comparison, along with the values in the Standard Model.}
\end{table}

We consider here three different lattice setups. The parameter sets for the standard Wilson action employed are given in table~\ref{tab:parameter_sets}. The properties and spectroscopy of these setups have been previously studied in detail in~\cite{Jenny:2022atm,Wurtz:2013ova,Maas:2026dlq}, for which we refer for details, including for the generation of configurations. As indicated by the tree-level parameters, all our sets are weakly coupled, but the coupling is stronger than in the Standard Model. We therefore expect that splitting in the Standard Model would be weaker, which may affect the PDFs and slow down the evolution from~\prefr{tlpdf1}{tlpdf4} towards higher energies.

The parameters of the first set H are taken from~\cite{Jenny:2022atm,Maas:2014pba}, and describe a fairly heavy Higgs boson, close but below the elastic threshold. The additional measurement of the phase shift in the scalar channel determined in~\cite{Jenny:2022atm} supplies information about the scattering length and thus an effective size parameter of the Higgs. Its value is about 40 GeV$^{-1}$; comparable to the value found for the physical vector bosons in~\cite{Maas:2018ska}, providing additional information on the bound state structure. Note, that this study was conducted at a relatively coarse lattice spacing, with a heavy Higgs mass at 1.85$m_W$ close to the elastic decay threshold and relatively strong couplings. 

The second set L from~\cite{Wurtz:2013ova} has a significantly finer lattice spacing with a Standard-Model-like Higgs-W mass hierarchy with weaker couplings aimed at more Standard-Model-like behavior. However, the vacuum expectation value is significantly lower than in the Standard Model, corresponding to stronger coupling causing substantial renormalization effects. Comparing its result to H will allow us to understand the sensitivity to the parameters of the theory, as well as to the discretization.

The final set F is exploratory. It adds two generations of dynamical fermions to set L. More details about this set are given in~\cite{Maas:2026dlq}. The fermions are still fairly heavy, a concession to the substantial amount of statistics necessary to evaluate PDFs. Still, their mass is comparable to the Higgs and the vector bosons. As they appear not as valence particles for the states considered here, and we do not yet have tree-level Yukawa couplings, we expected their presence to produce only a negligible effect, which is confirmed by the numerical results.

\begin{table}
    \centering
    \begin{tabular}{|c|c|c|c|} 
    \hline
         $N$ & H & L & F  \cr
         \hline
         8  & 972947  &         &  \cr
         12 & 436890  & 404193 & \cr
         16 & 399404  & 332568 & 37771 \cr
         20 & 207920  & 227103 & 19900 \cr
         24 & 199550  & 216280 & \cr
         28 & 119353  & 139495 & \cr
         32 & 96839   &        & \cr
         \hline
    \end{tabular}
    \caption{\label{tab:statistics}Number of configurations used for each parameter set and lattice size.}
\end{table}

We list the number of decorrelated configurations in table~\ref{tab:statistics}, where we used the plaquette to measure residual autocorrelations. The relevant observables are three-point functions of composite operators which are fairly noisy for this theory~\cite{Maas:2018ska} but also in general, see QCD applications~\cite{Cichy:2018mum}. As a result, a substantial number of configurations was needed. This restricted the maximum lattice size and mass of the fermions for set F. Fortunately, as will be seen, the results are fairly independent of volume, in accordance with other investigations of this theory~\cite{Maas:2014pba,Afferrante:2020fhd}, such that relatively small volumes are enough to suppress systematic volume effects sufficiently for the present purpose.

\subsection{Matrix element determination}

For the involved operators,~\prefr{op1}{op2} and~\prefr{f0}{f5}, the Wilson line and field strength tensors in~\prefr{f0}{f5} are both expressed in terms of products of links $U_\mu$ along the shortest straight connecting path between both points. The choice follows for the operators~\cite{Evertz:1985fc,Maas:2014pba} and for the currents~\cite{Constantinou:2020hdm,HadStruc:2021wmh,Balitsky:2019krf,Good:2024iur}. This yields
\bea
O_0(x)&=&\det X(x)\,,\nn\\
O_1^{a\mu}(x)&=&\tr \left(\tau^a\frac{X(x)^\dagger}{\sqrt{\det X(x)^\dagger}}U_\mu(x)\frac{X(x+\mu)^\dagger}{\sqrt{\det X(x+\mu)^\dagger}}\right)\,,\nn\\
W({\cal C}_{z\to z_0})&=&\left(U_3(z_0)\times...\times U_3(z)\right)^\dagger\,,\nn\\
W_\mn(x)&=&\frac{i}{8}\left(U_\mn+U_{\nu-\mu}+U_{-\mu-\nu}+U_{-\nu\mu}-U_{\nu\mu}-U_{-\mu\nu}-U_{-\nu-\mu}-U_{\mu-\nu}\right)
\eea
\no for the necessary lattice versions of the operators, where $\mu$ is a lattice unit vector in $\mu$ direction, $U_\mu$ a link, and $U_\mn$ the plaquette. Taking only the SU(2) part of the Higgs field (this refers to the additional division by the determinant instead of a trivial implementation of equation~\pref{op2}) for $O_1$ has been found to substantially reduce fluctuations without changing the overlap with the asymptotic state~\cite{Evertz:1985fc}. The Wilson line $W$ is now taken along a straight line connecting points differing only in their 3-direction, ranging from $z$ to $z_0$, and determined in the usual way~\cite{Montvay:1994cy} by a product of links variables.

To further enhance the signal, we smeared the bosonic operators using up to 4 level of APE smearing~\cite{Philipsen:1996af,Maas:2014pba} as detailed in~\cite{Maas:2026dlq}. Note that on our smallest lattice of $8^4$ this implies that smeared sinks and sources will overlap with the probe current, which will be discussed as a finite volume effect corresponding to keeping a fixed smearing radius relative to the lattice size, which will become point-like in the thermodynamic limit. We transform to momentum space using a standard discrete Fourier transform. Finally, the integral in~\pref{mfunc} is computed by summing over $0<\tau<t$ followed by a numerical derivative. To obtain the matrix element, we fit the plateau of the resulting quantity. The quasi-PDF~\pref{quasipdf} is determined by the discrete sum over the possible values of $z$ from the lattice. We further average over all possible axes for the choice of the straight path from $z$ to $z_0$. 

We found that, while the function $M(P_3,z)$  should decay for large $z$ to zero, volume artifacts lead to a non-zero value causing an oscillatory quasi-PDF behavior which persists at $P_3\to\infty$ and is not restricted to the interval $x\in[0,1]$. This constant contribution decreases for increasing $P_3$, but the decrease is not sufficiently fast to compensate for the increase in statistical fluctuations at higher momenta. We thus explicitly subtracted this constant from the function $M(P_3,z)$, eliminating this finite volume artifact. This created a fairly smooth behavior in volume. In practice, we determine the constant by averaging the two largest $z$ values to determine the constant to be subtracted, which reduces the impact of statistical fluctuations. As this is not an exact procedure, a small oscillatory contribution remains.

Statistical errors are computed on the matrix-element level. Our primary matrix elements have very small errors, but the signal deteriorates exponentially with both $P_3$ and $t$. They are propagated to the quasi-PDF by varying the central values by $\pm1\sigma$ before determining the quasi-PDF. The errors at different values of $x$ are thus highly correlated due to the various averaging and fitting steps in between, especially in~\pref{mfunc} and~\pref{quasipdf}. The final error band is therefore not representing a situation where the error for each $x$ value is an independent statistical fluctuation. Rather, the propagation of the errors in the numerical process smooths out the fluctuations, giving the continuous error band shown, which is highly correlated between different $x$ values. Eventually, we take the Bjorken limit to be the point where the quasi-PDF remains stable within errors.

\section{Lattice results}\label{s:lpdf}

A study along different lines of constant physics is not strictly necessary at the current level of precision, since due to the weak coupling, the scale dependence is expected to be weak. This notion is corroborated by our data: The spectrum as well as form factors showed no strong dependence on the lattice spacing~\cite{Maas:2014pba,Maas:2018ska}. Given the lack of excited states in the spectrum below the relevant inelastic thresholds~\cite{Jenny:2022atm,Maas:2014pba,Wurtz:2013ova}, combined with a light ground state spectrum, this is not entirely surprising. 

This leaves two major sources of systematic uncertainties, the volume and the extrapolation of $P_3\to\infty$ for the final result. Both are not entirely disjoint. On small volumes only a smaller range of $P_3$ is available, which, however, already corresponds to much larger momenta and thus has higher statistical errors.

In total, we investigated 252 quasi-PDFs as a function of the lattice momentum $P_3$ on different volumes, obtained from several millions of individual matrix element measurements for the different values of $z$ ranging from $-N/2$ to $N/2$, $t=3,\dots,N/2$, $\tau=1,\dots,t$, and $P_3=0,\pi/N,\dots,\pi$. The results for the matrix elements are very stable, justifying the correctness of our estimation of uncertainties. The matrix elements are made available in the dedicated data release~\cite{data_release} for reproducibility. This paper will only show the results on the final quasi-PDFs.

\subsection{Volume effects}

\begin{figure}
 \includegraphics[width=0.5\textwidth]{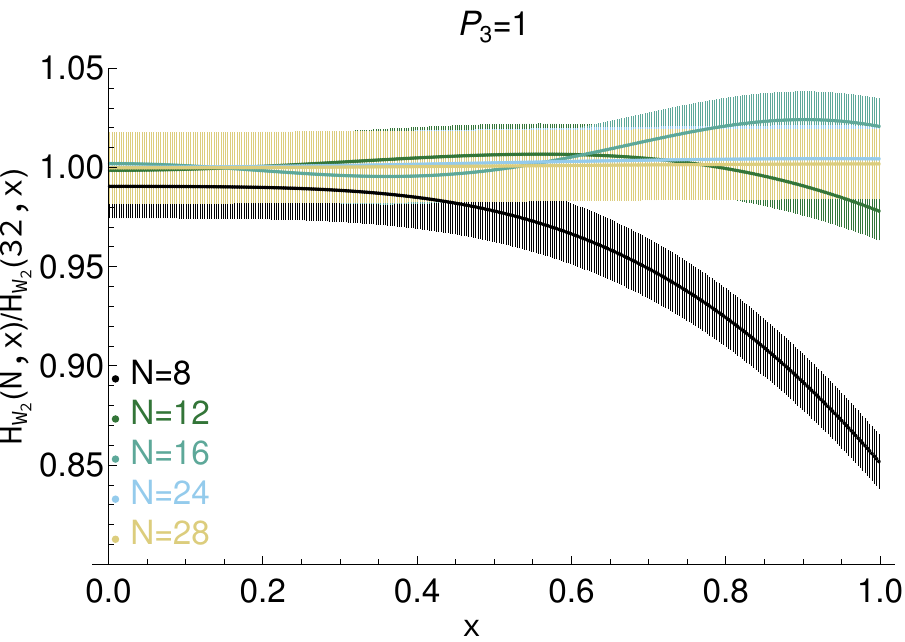}\includegraphics[width=0.5\textwidth]{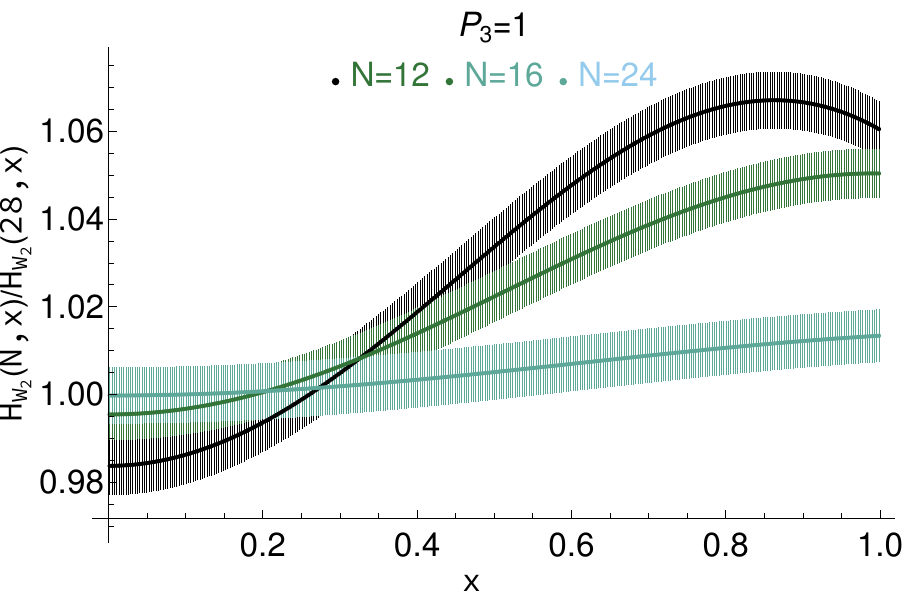}\\
 \includegraphics[width=0.5\textwidth]{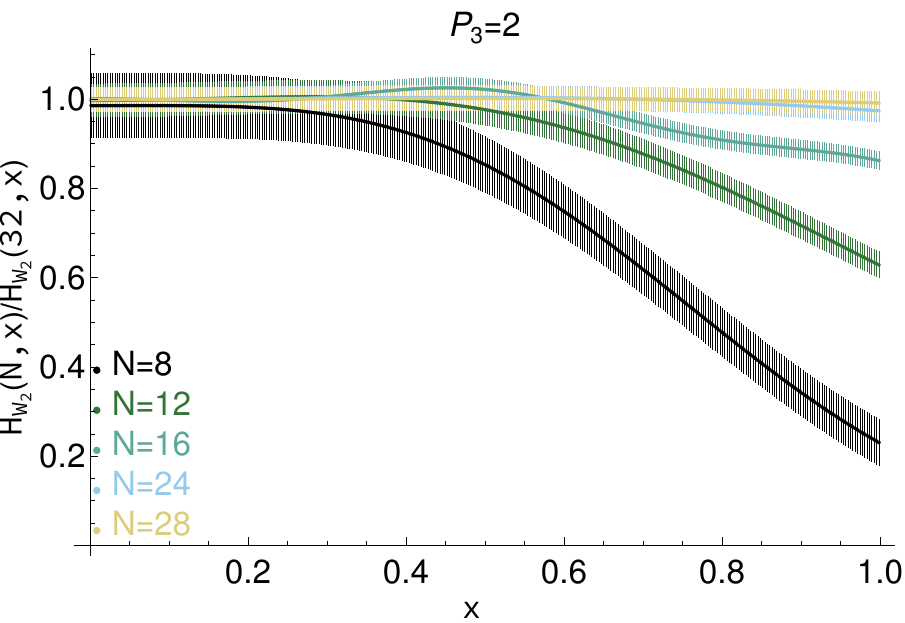}\includegraphics[width=0.5\textwidth]{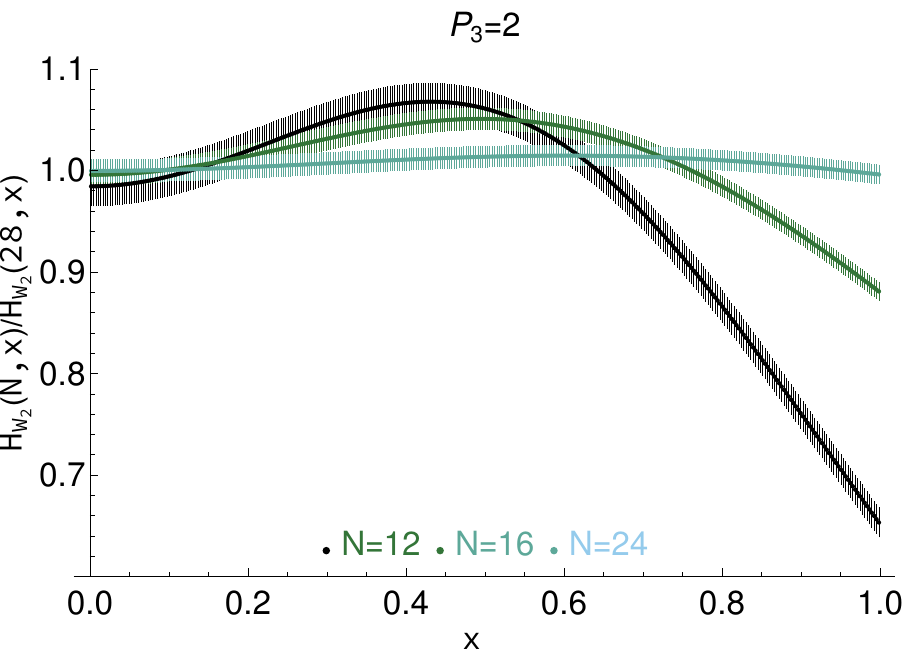}\\
 \includegraphics[width=0.5\textwidth]{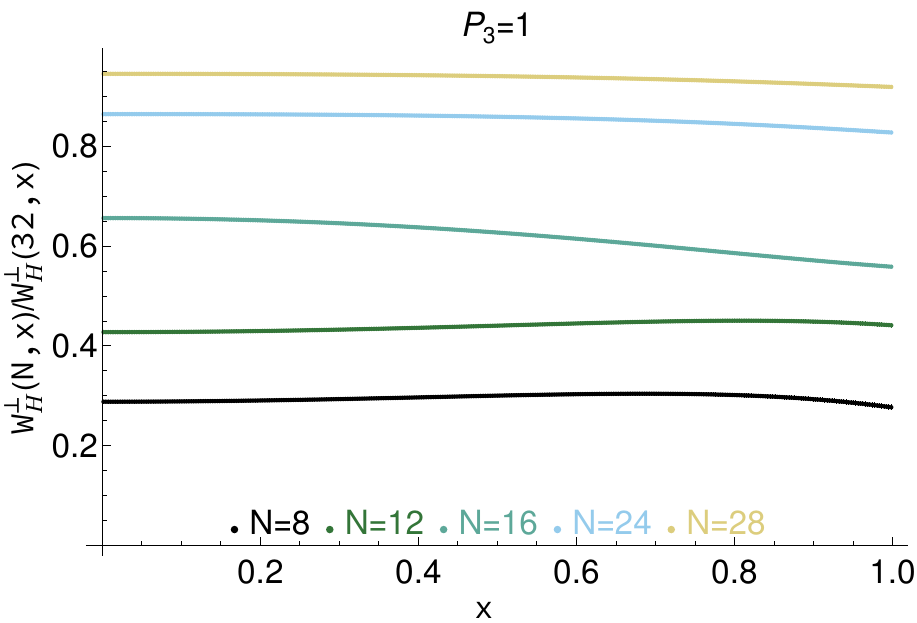}\includegraphics[width=0.5\textwidth]{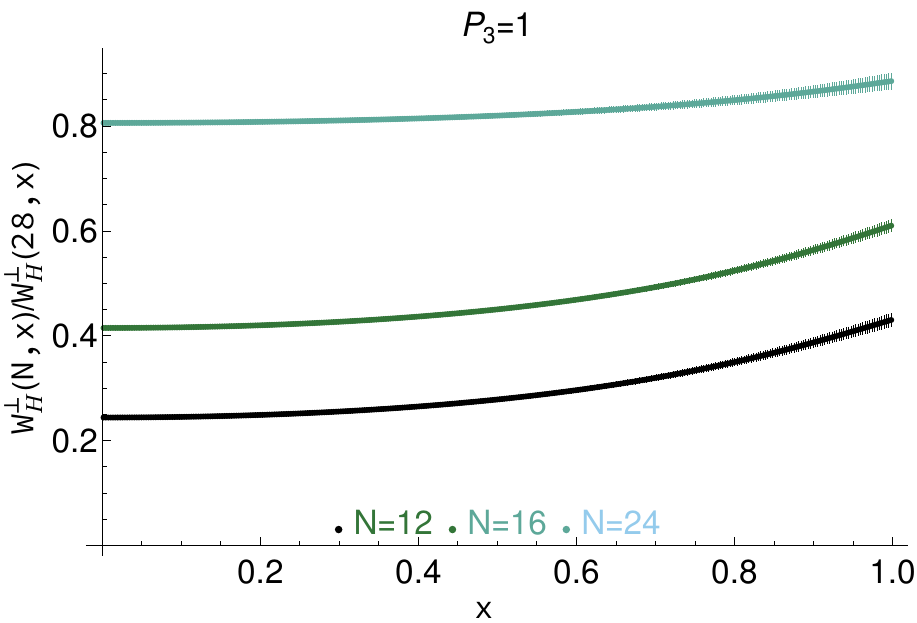}\\
 \includegraphics[width=0.5\textwidth]{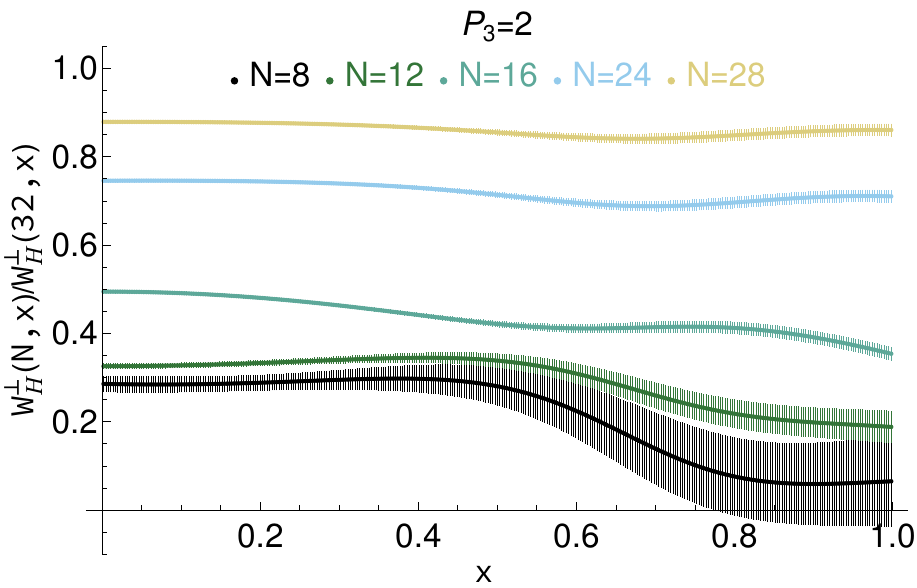}\includegraphics[width=0.5\textwidth]{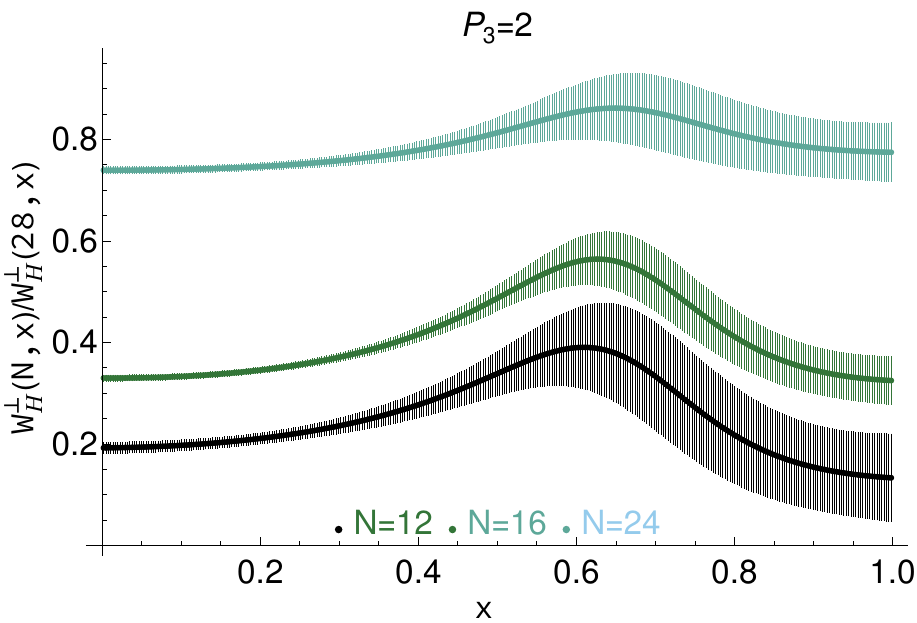}
 \caption{The ratio of the vector quasi-PDFs $H_{W_2}$ of the scalar (top panels) and the scalar quasi-PDFs $W^\perp_{H}$ of the transverse polarized vector (bottom panels) as ratios between fixed volume to the largest volumes, and as a function of $x$. The topmost row and the third row are for the lowest non-vanishing momentum and the second-to-top row and the bottom row for the second-lowest non-vanishing momentum. The left column is from the H ensemble and the right column from the L ensemble.}
 \label{vol}
\end{figure}

We measure the impact of the final volume effects by studying the ratio of the extracted quasi-PDF on a given volume to the one on the maximal volume. For all sets and all currents, we find that this ratio quickly approaches a constant when increasing the volume. To illustrate this, examples are shown in figure~\ref{vol}. Even for fairly small volumes the ratio is almost constant within a few percent. Furthermore, the differences are mainly constrained to the largest $x$-values. This is true for all momenta. However, due to the quick increase of noise when decreasing the volume and increasing the lattice momentum, this statement becomes less statistical reliable towards the Bjorken limit. Still, for the largest volumes the ratio remain fairly constant within the larger errors.    

Figure~\ref{vol} shows the scalar quasi-PDF for the transversely polarized vector $W^\perp_H$  and the vector current for the scalar $H_{W_2}$. In the latter case, we use the current~\pref{f2}. Nevertheless, all vector currents with a statistically significant signal give the same qualitative result. There, it is directly visible that for the second-largest volumes the ratio to the largest volume is nearly $x$-independent within a few percent, though the ratio itself is not necessarily one. The latter effect will be absorbed in the normalization. In fact, very often already volumes as small as $16^4$ give a reasonably flat ratio.

This pattern repeats itself for every quasi-PDF, operator, and set. Thus, we conclude that volume effects are essentially negligible for our largest volumes. We will therefore in the following only consider this largest volume. The only one, where some finite volume effects are still relevant at large $x$ is the computationally most expensive set F. However, the final results for the ensemble F are very close to the results of the pure gauge sets H and L. But our primary observation, that dynamical heavy fermions have negligible impact on the studied quasi-PDFs, is not affected by this. Hence, also in this case we will use in the following only the largest volume.

\subsection{Fits for $P_3\to\infty$}

The results from the lattice for the quasi-PDFs depend on the momentum $P_3$. The Bjorken limit is defined as the limit $P_3\to\infty$. To get an estimate of this limit, the quasi-PDFs were fitted for each $P_3$ independently, followed by an extrapolation. 

For the fits of the $x$-dependence we used the following ans{\"a}tze
\bea
f(x)&=&\frac{a_1}{a_2+x^\kappa_1}\label{fit1}\,,\\
f(x)&=&\frac{a_1}{x}\left(1-a_2 x^{\kappa_1}\right)\label{fit2}\,,\\
f(x)&=&\frac{1}{x}(1-x)^{\kappa_1}(a_1+a_2 x^2)\label{fit3}\,,\\
f(x)&=&\frac{1}{x}\frac{1}{a_2+\frac{x^\kappa_1}{a_1}}\label{fit4}\,,
\eea
\no which where sufficient to describe all PDFs. In particular, for any given PDF, the same fit form was working for all $P_3$. However, the $P_3$ dependence of the parameters $a_1$, $a_2$, and $\kappa_1$ in~\prefr{fit1}{fit4} was more variable. To extrapolate to the Bjorken limit, we fitted these parameters with the following fit functions
\bea
p(P_3)&=&a+\frac{b}{P_3^2}\label{pfit1}\,,\\
p(P_3)&=&a+\frac{b}{P_3}\label{pfit1a}\,,\\
p(P_3)&=&a+\frac{b}{P_3^c}\label{pfit2}\,,\\
p(P_3)&=&a^2\label{pfit3}\,,\\
p(P_3)&=&\theta(P_3-3.5)\left(1-\frac{b}{P_3^\frac{1}{c}}\right)\label{pfit4}\,,\\
p(P_3)&=&\theta(P_3-4.5)\left(1-\frac{b}{P_3^\frac{1}{c}}\right)\label{pfit4a}\,,\\
p(P_3)&=&\theta(3.5-P_3)\left(a+\frac{b}{3.5-P_3}\right)+\theta(P_3-3.5)\left(c+\frac{d}{3.5-P_3}\right)\label{pfit5}\,,\\
p(P_3)&=&\theta(3.5-P_3)\left(a+\frac{b}{3.5-P_3}\right)+\theta(P_3-3.5)\left(c+\frac{d}{3.5-P_3^2}\right)\label{pfit5a}\,,\\
p(P_3)&=&\theta(4.5-P_3)\left(a+\frac{b}{4.5-P_3}\right)+\theta(P_3-4.5)\left(c+\frac{d}{4.5-P_3^2}\right)\label{pfit5b}\,,\\
p(P_3)&=&\frac{a}{2}\left(1+\tanh\left(\frac{P_3-b}{c}\right)\right)\label{pfit6}\,,\\
p(P_3)&=&d+\frac{a}{2}\left(1+\tanh\left(\frac{P_3-b}{c}\right)\right)\label{pfit6a}\,,\\
p(P_3)&=&1+\frac{b}{(1+c P_3)^d}\label{pfit7}\,,\\
p(P_3)&=&\frac{a}{b+P_3^c}\label{pfit8}\,.
\eea
\no The appearance of the $\theta$-function in some of these has to do with the fact that some features only start to become statistically non-zero for sufficiently large values of $P_3$. However, using the fit forms at large $P_3$ also for small $P_3$ lead to no stable fits. Therefore, we use here this solution with step-wise fit forms. With this ansatz, the fit parameters $a_1$, $a_2$, and $\kappa_1$ for each PDF become, within statistical uncertainties, $P_3$-independent for sufficiently large values of $P_3$. This strongly suggests that we indeed can extrapolate to the Bjorken limit in this way. 
\begin{table}
    \centering
    \begin{tabular}{|c|c|c|c|c|c|c|} 
    \hline
         Set & PDF & Fit & $\max P_3$ & $a_1$ & $a_2$ & $\kappa_1$ \cr
         \hline
         \hline
         H & $H_H$ &~\pref{fit1} & 5 & $0.0014^{+8}_{-6}$~\pref{pfit1} & $0.012^{+16}_{-11}$~\pref{pfit2} & $2.1^{+3}_{-2}$~\pref{pfit1} \cr
         \hline
         H & $H_{W_1}$ &~\pref{fit2} & 6 & $0.213^{+10}_{-11}$~\pref{pfit3} & 1~\pref{pfit4} & $3.4^{+4}_{-3}$~\pref{pfit5} \cr
         \hline
         H & $H_{W_2}$ &~\pref{fit2} & 6 & $0.261^{+12}_{-13}$~\pref{pfit3} & 1~\pref{pfit4} & $3.3^{+4}_{-3}$~\pref{pfit5} \cr
         \hline
         H & $H_{W_3}$ &~\pref{fit3} & 6 & $13^{+34}_{-11}\times 10^{-5}$~\pref{pfit6} & $0.0002^{+15}_{-1}$~\pref{pfit6} & $2.8(1)$~\pref{pfit5} \cr
         \hline
         H & $H_{W_4}$ &~\pref{fit2} & 7 & $0.248^{+15}_{-16}$~\pref{pfit3} & $1.0(7)$~\pref{pfit6} & 1~\pref{pfit7} \cr
         \hline
         H & $H_{W_5}$ &~\pref{fit2} & 7 & $0.176(11)$~\pref{pfit3} & $1.1(8)$~\pref{pfit6} & 1~\pref{pfit7} \cr 
         \hline
         \hline
         H & $W_H^\parallel$ &~\pref{fit1} & 6 & $0.0012^{+3}_{-2}$~\pref{pfit1} & $0.010^{+5}_{-4}$~\pref{pfit2} & $2.10^{+9}_{-8}$~\pref{pfit1} \cr
         \hline
         H & $W_{W_1}^\parallel$ &~\pref{fit2} & 9 & $0.039(3)$~\pref{pfit1} & $1.1^{+6}_{-3}$~\pref{pfit6} & $1.8^{+2}_{-1}$~\pref{pfit5a} \cr
         \hline
         H & $W_{W_2}^\parallel$ &~\pref{fit2} & 9 & $0.058(6)$~\pref{pfit1} & $1.1^{+6}_{-3}$~\pref{pfit6} & $1.83^{+11}_{-10}$~\pref{pfit5a} \cr
         \hline
         H & $W_{W_3}^\parallel$ & - & 9 & - & - & - \cr
         \hline
         H & $W_{W_4}^\parallel$ &~\pref{fit2} & 9 & $0.053(4)$~\pref{pfit2} & $1.2^{+4}_{-2}$~\pref{pfit6} & 1~\pref{pfit7} \cr
         \hline
         H & $W_{W_5}^\parallel$ &~\pref{fit2} & 9 & $0.026(2)$~\pref{pfit2} & $1.2^{+4}_{-2}$~\pref{pfit6} & $0.90(1)$~\pref{pfit6a} \cr
         \hline
         \hline
         H & $W_H^\perp$ &~\pref{fit1} & 6 & $0.00007^{+7}_{-5}$~\pref{pfit1} & $0.010^{+8}_{-7}$~\pref{pfit2} & $2.1(2)$~\pref{pfit1} \cr 
         \hline
         H & $W_{W_1}^\perp$ &~\pref{fit2} & 8 & $0.0023(7)$~\pref{pfit1a} & $1.0^{+9}_{-3}$~\pref{pfit6} & $2.09(8)$~\pref{pfit5a} \cr
         \hline
         H & $W_{W_2}^\perp$ &~\pref{fit2} & 8 & $0.0035(11)$~\pref{pfit1a} & $1.0^{+8}_{-3}$~\pref{pfit6} & $2.10(8)$~\pref{pfit5a} \cr
         \hline
         H & $W_{W_3}^\perp$ & - & 8 & - & - & - \cr
         \hline
         H & $W_{W_4}^\perp$ &~\pref{fit2} & 8 & 0.0032(8)~\pref{pfit1a} & $1.1^{+6}_{-3}$~\pref{pfit6} & 1~\pref{pfit7} \cr
         \hline
         H & $W_{W_5}^\perp$ &~\pref{fit2} & 8 & 0.0032(8)~\pref{pfit1a} & $1.1^{+6}_{-3}$~\pref{pfit6} & 1~\pref{pfit7} \cr
         \hline
         \hline
    \end{tabular}
    \caption{\label{tab:pdffitH} Unnormalized fits of the raw PDFs for set H. Fit refers to the fit form~\prefr{fit1}{fit4} used to fit the PDF itself, and - implies none was possible. The references in parentheses for the fit parameters refers to which extrapolation form~\prefr{pfit1}{pfit8} has been used for the determination of the $P_3\to\infty$ of the parameters. Note that a positivity cut has been applied to all fitted parameters.}
\end{table}

\begin{table}
    \centering
    \begin{tabular}{|c|c|c|c|c|c|c|} 
    \hline
         Set & PDF & Fit & $\max P_3$ & $a_1$ & $a_2$ & $\kappa_1$ \cr
         \hline
         \hline
         L & $H_H$ &~\pref{fit1} & 4 & $0.0005^{+13}_{-5}$~\pref{pfit1} & $0.008^{+12}_{-8}$~\pref{pfit2} & $2.6^{+6}_{-5}$~\pref{pfit1} \cr
         \hline
         L & $H_{W_1}$ &~\pref{fit2} & 7 & $0.085(5)$~\pref{pfit3} & 1~\pref{pfit4} & $0.9^{+7}_{-5}$~\pref{pfit5b} \cr
         \hline
         L & $H_{W_2}$ &~\pref{fit2} & 7 & $0.104(6)$~\pref{pfit3} & 1~\pref{pfit4} & $0.7^{+7}_{-5}$~\pref{pfit5b} \cr
         \hline
         L & $H_{W_3}$ & - & 7 & - & - & - \cr
         \hline
         L & $H_{W_4}$ &~\pref{fit4} & 6 & $0.1^{30}_{1}\times 10^{-6}$~\pref{pfit1} & $9.1(2)$~\pref{pfit3} & $2.10^{+4}_{-3}$~\pref{pfit1} \cr
         \hline
         L & $H_{W_5}$ &~\pref{fit4} & 6 & $10^{+8}_{-10}\times 10^{-6}$~\pref{pfit1} & $12.9(2)$~\pref{fit3} & $2.1^{+4}_{-3}$~\pref{pfit1} \cr
         \hline
         \hline
         L & $W_H^\parallel$ &~\pref{fit1} & 4 & $0.0006^{+5}_{-3}$~\pref{pfit1} & $0.007^{+9}_{-6}$~\pref{pfit2} & $2.6(3)$~\pref{pfit1} \cr
         \hline
         L & $W_{W_1}^\parallel$ &~\pref{fit2} & 8 & $0.0066(5)$~\pref{pfit1} & 1~\pref{pfit4a} & $1.1(2)$~\pref{pfit5b} \cr
         \hline
         L & $W_{W_2}^\parallel$ &~\pref{fit2} & 8 & $0.0099(8)$~\pref{pfit1} & 1~\pref{pfit4a} & 1.1(2)~\pref{pfit5b} \cr
         \hline
         L & $W_{W_3}^\parallel$ & - & 8 & - & - & - \cr
         \hline
         L & $W_{W_4}^\parallel$ &~\pref{fit4} & 8 & $0.00003(3)$~\pref{pfit2} & $85^{+10}_{-13}$~\pref{fit2} & $2.15^{+6}_{-5}$~\pref{pfit1} \cr
         \hline
         L & $W_{W_5}^\parallel$ &~\pref{fit4} & 8 & $2^{+7}_{-2}\times 10^{-6}$~\pref{pfit2} & $178^{+10}_{-8}$~\pref{pfit1} & $2.12^{+6}_{-5}$~\pref{pfit1} \cr
         \hline
         \hline
         L & $W_H^\perp$ &~\pref{fit1} & 4 & $4^{+8}_{-4}\times 10^{-5}$~\pref{fit2} & $0.007^{+9}_{-6}$~\pref{pfit2} & $2.6(3)$~\pref{pfit1} \cr
         \hline
         L & $W_{W_1}^\perp$ &~\pref{fit2} & 8 & $0.00^{+9}_{-0}$~\pref{pfit2} & 1~\pref{pfit4a} & $1.0(2)$~\pref{pfit5b} \cr
         \hline
         L & $W_{W_2}^\perp$ &~\pref{fit2} & 8 & $0.00^{+9}_{-0}$~\pref{pfit2} & 1~\pref{pfit4a} & $0.9(2)$~\pref{pfit5b} \cr
         \hline
         L & $W_{W_3}^\perp$ & - & 9 & - & - & - \cr
         \hline
         L & $W_{W_4}^\perp$ &~\pref{fit4} & 7 & $2^{+10}_{-2}\times 10^{-5}$~\pref{pfit2} & $118^{+8}_{-6}$~\pref{pfit8} & $2.14(4)$~\pref{pfit1} \cr
         \hline
         L & $W_{W_5}^\perp$ &~\pref{fit4} & 7 & $2^{+11}_{-2}\times 10^{-5}$~\pref{pfit2} & $118^{+8}_{-6}$~\pref{pfit8} & $2.14(4)$~\pref{pfit1} \cr
         \hline
         \hline
    \end{tabular}
    \caption{\label{tab:pdffitL} Unnormalized fits of the raw PDFs for set L. Fit refers to the fit form~\prefr{fit1}{fit4} used to fit the PDF itself, and - implies none was possible. The references in parentheses for the fit parameters refers to which extrapolation form~\prefr{pfit1}{pfit8} has been used for the determination of the $P_3\to\infty$ of the parameters. Note that a positivity cut has been applied to all fitted parameters.}
\end{table}

\begin{table}
    \centering
    \begin{tabular}{|c|c|c|c|c|c|c|} 
    \hline
         Set & PDF & Fit & $\max P_3$ & $a_1$ & $a_2$ & $\kappa_1$ \cr
         \hline
         \hline
         F & $H_H$ &~\pref{fit1} & 3 & $0.0004^{+87}_{-4}$~\pref{pfit1} & $0.00^{+14}_{-0}$~\pref{pfit1} & $3^{+4}_{-2}$~\pref{pfit1} \cr
         \hline
         F & $H_{W_1}$ &~\pref{fit2} & 4 & $0.084^{+4}_{-5}$~\pref{pfit3} & 1~\pref{pfit4} & $3.1(3)$~\pref{pfit6} \cr
         \hline
         F & $H_{W_2}$ &~\pref{fit2} & 4 & $0.103^{+5}_{-6}$~\pref{pfit3} & 1~\pref{pfit4} & $3.0(3)$~\pref{pfit6} \cr
         \hline
         F & $H_{W_3}$ & - & 4 & - & - & - \cr
         \hline
         F & $H_{W_4}$ &~\pref{fit4} & 4 & $0.000^{+2}_{-0}$~\pref{pfit1} & $9.2^{+4}_{-3}$~\pref{pfit3} & $2.11^{+9}_{-7}$~\pref{pfit1} \cr
         \hline
         F & $H_{W_5}$ &~\pref{fit4} & 4 & $0.000^{+7}_{-0}$~\pref{pfit1} & $12.9^{+5}_{-4}$~\pref{pfit3} & $2.12^{+9}{-7}$~\pref{pfit1} \cr\hline
         \hline
         \hline
         F & $W^\parallel_{H}$ &~\pref{fit1} & 3 & $0.001^{+3}_{-1}$~\pref{pfit1} & $0.00^{+7}_{-0}$~\pref{pfit1} & $2.7^{+10}_{-5}$~\pref{pfit3} \cr
         \hline
         F & $W^\parallel_{W_1}$ &~\pref{fit2} & 5 & $0.007(2)$~\pref{pfit1} & 1~\pref{pfit4} & $2.5(4)$~\pref{pfit5} \cr
         \hline
         F & $W^\parallel_{W_2}$ &~\pref{fit2} & 5 & $0.010(2)$~\pref{pfit1} & 1~\pref{pfit4} & $2.4^{+4}_{-3}$~\pref{pfit5} \cr
         \hline
         F & $W^\parallel_{W_3}$ & - & 5 & - & - & - \cr
         \hline
         F & $W^\parallel_{W_4}$ &~\pref{fit4} & 5 & $0.00^{+4}_{-0}$~\pref{pfit2} & $88^{+14}_{-10}$~\pref{pfit1} & $2.1^{+2}_{-1}$~\pref{pfit1} \cr
         \hline
         F & $W^\parallel_{W_5}$ &~\pref{fit4} & 5 & $0.000^{+2}_{-0}$~\pref{pfit2} & $177^{+30}_{-20}$~\pref{pfit1} & $2.2^{+3}_{-2}$~\pref{pfit1} \cr
         \hline
         \hline
         F & $W^\perp_{H}$ &~\pref{fit1} & 3 & $0.0000^{+4}_{-0}$~\pref{pfit1} & $0.00^{+7}_{-0}~\pref{pfit1}$ & $2.7^{+10}_{-5}$~\pref{pfit3} \cr
         \hline
         F & $W^\perp_{W_1}$ &~\pref{fit2} & 5 & $0.00007^{+43}_{-7}$~\pref{pfit2} & 1~\pref{pfit4a} & $0.9992^{+4}_{-9}$~\pref{pfit3} \cr
         \hline
         F & $W^\perp_{W_2}$ &~\pref{fit2} & 5 & $0.0001^{+5}_{-1}$~\pref{pfit2} & 1~\pref{pfit4} & $2.3^{+4}_{-3}$~\pref{pfit5} \cr
         \hline
         F & $W^\perp_{W_3}$ &~\pref{fit3} & 5 & $0.00020(6)$~\pref{pfit3} & $0.00008^{+9}_{-4}$~\pref{pfit3} & $80^{+50}_{-80}$~\pref{pfit6} \cr
         \hline
         F & $W^\perp_{W_4}$ &~\pref{fit4} & 5 & $0.0000^{+2}_{-0}$~\pref{pfit2} & $225^{+41}_{-122}$~\pref{pfit8} & $2.1^{+2}_{-7}$~\pref{pfit3} \cr
         \hline
         F & $W^\perp_{W_5}$ &~\pref{fit4} & 5 & $0.0000^{+2}_{-0}$~\pref{pfit3} & $224^{++41}_{-122}$~\pref{pfit8} & $2.1^{+2}_{-1}$~\pref{pfit3} \cr
         \hline
         \hline
    \end{tabular}
    \caption{\label{tab:pdffitF}Unnormalized fits of the raw PDFs for set F. Fit refers to the fit form~\prefr{fit1}{fit4} used to fit the PDF itself, and - implies none was possible. The references in parentheses for the fit parameters refers to which extrapolation form~\prefr{pfit1}{pfit8} has been used for the determination of the $P_3\to\infty$ of the parameters. Note that a positivity cut has been applied to all fitted parameters.}
\end{table}

We list the results for the parameters in the Bjorken limit from the fit functions~\prefr{fit1}{fit4} in tables~\ref{tab:pdffitH},~\ref{tab:pdffitL}, and~\ref{tab:pdffitF}, and indicate up to which value of $P_3$ the data was used to perform an extrapolation using the fit formulas~\prefr{pfit1}{pfit8}. To estimate the errors on the fitted parameters, we repeated the fits with the PDFs evaluated at the boundaries of their error bands. Given that our errors between different values of $x$ are strongly correlated, this appears justified, to get an estimate of the variability of the resulting Bjorken limit.

Some of the fit parameters in tables~\prefr{tab:pdffitH}{tab:pdffitF} are compatible with zero. This is an artifact of the missing renormalization paired with a large statistical noise. While the finite-momentum quasi-PDFs are always statistically significant non-zero, the uncertainties when extrapolating to infinite $P_3$ allows for a compatibility with zero. This was, however, irrelevant for the final results in section~\ref{s:pdf}, as the normalization is entirely determined by the sum rules of section~\ref{s:normalization}.

\begin{figure}
 \includegraphics[width=0.5\textwidth]{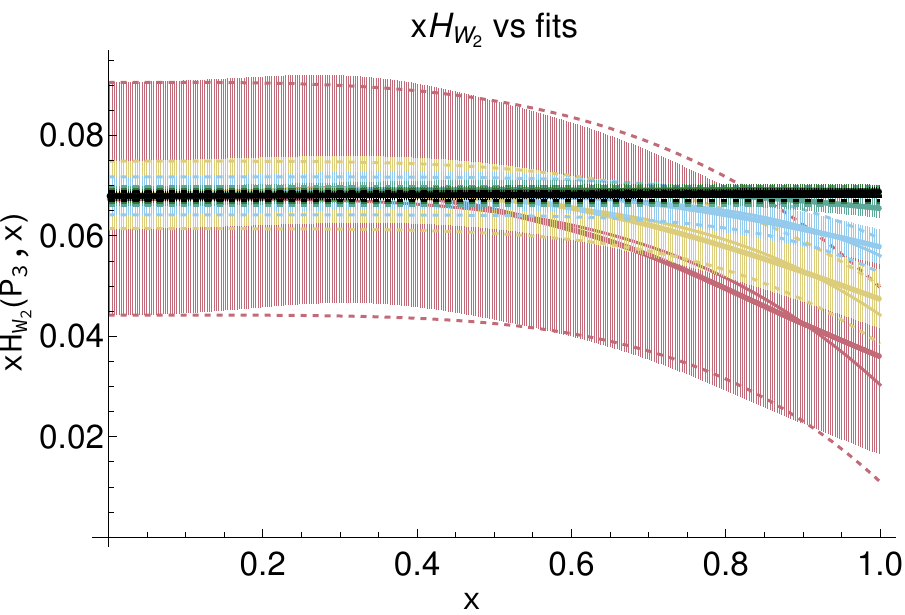}\includegraphics[width=0.5\textwidth]{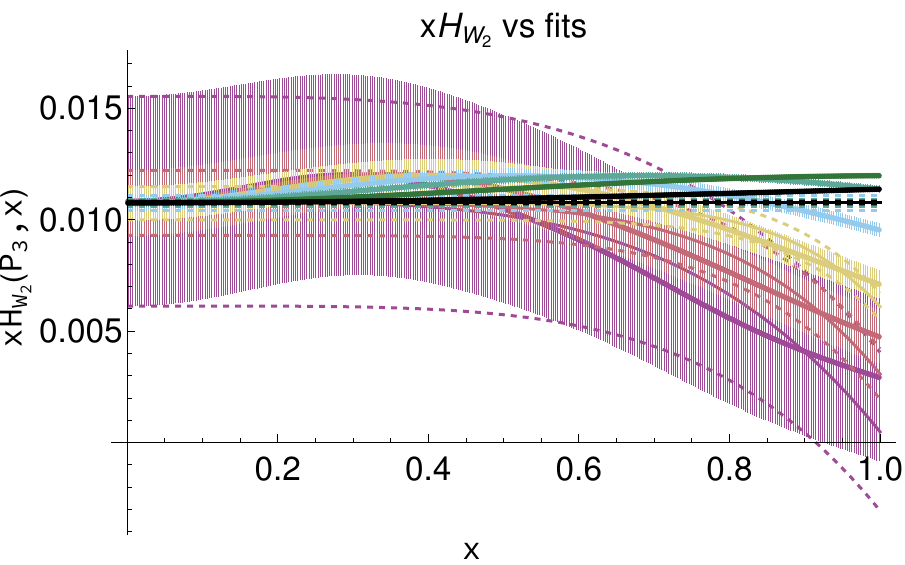}\\
 \includegraphics[width=0.5\textwidth]{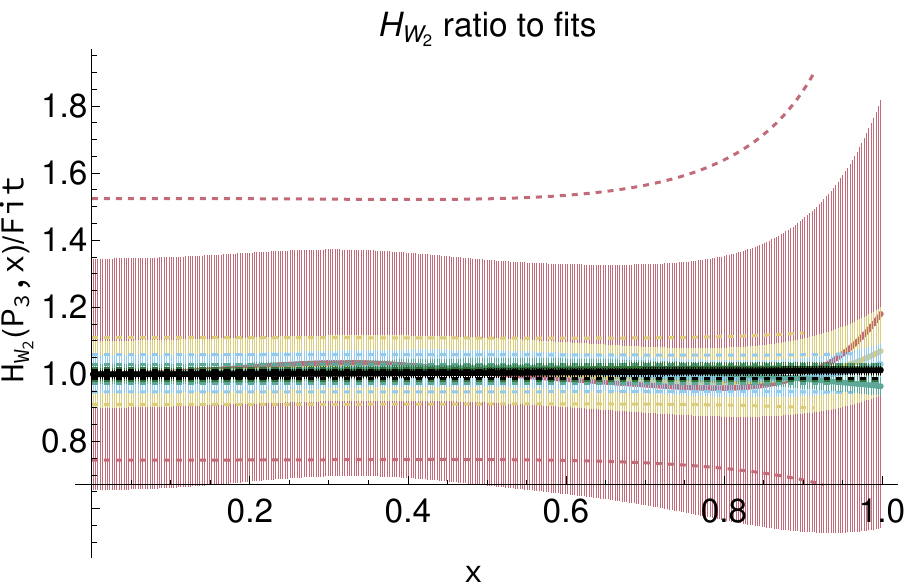}\includegraphics[width=0.5\textwidth]{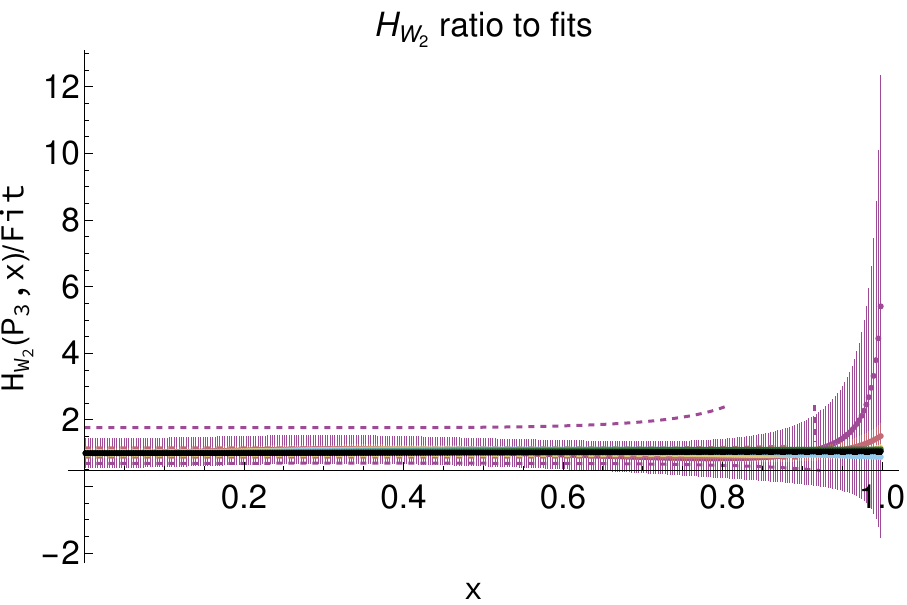}\\
 \includegraphics[width=0.5\textwidth]{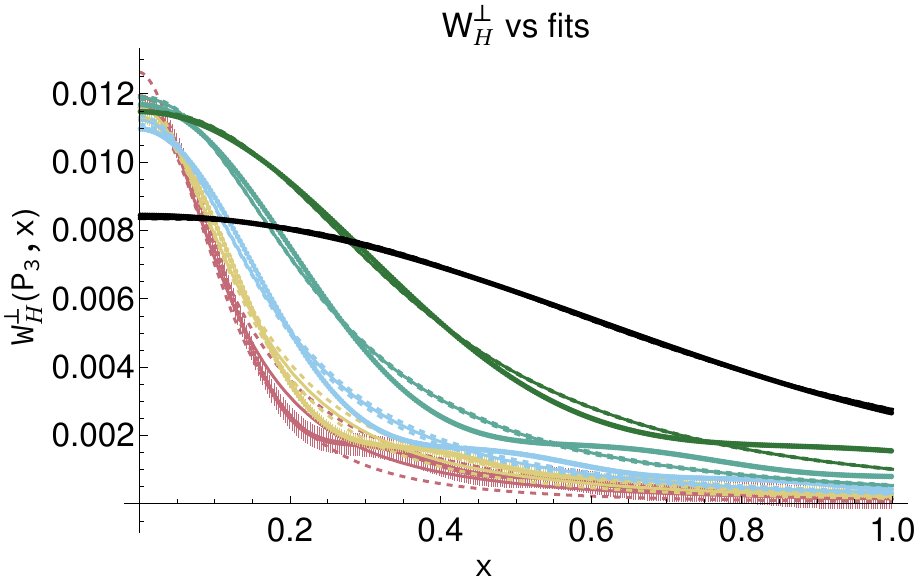}\includegraphics[width=0.5\textwidth]{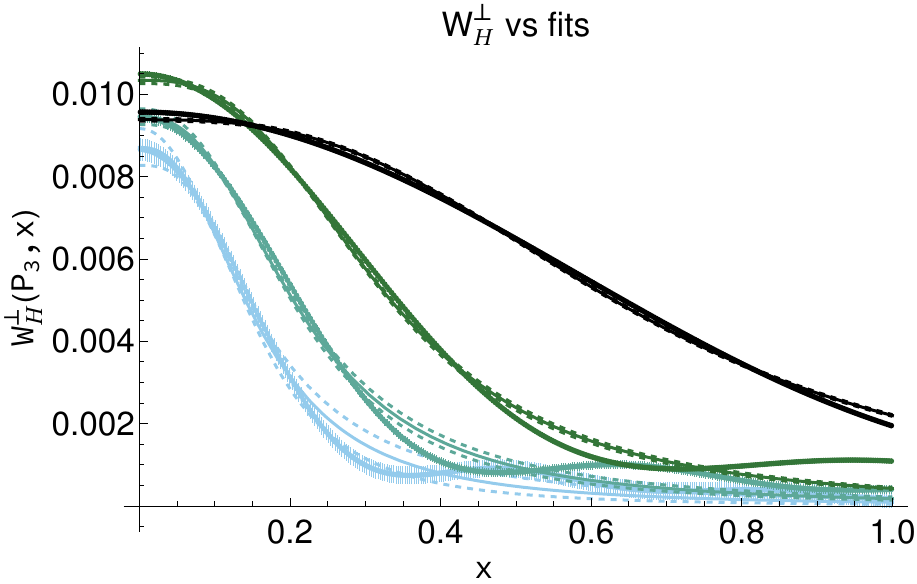}\\
 \includegraphics[width=0.5\textwidth]{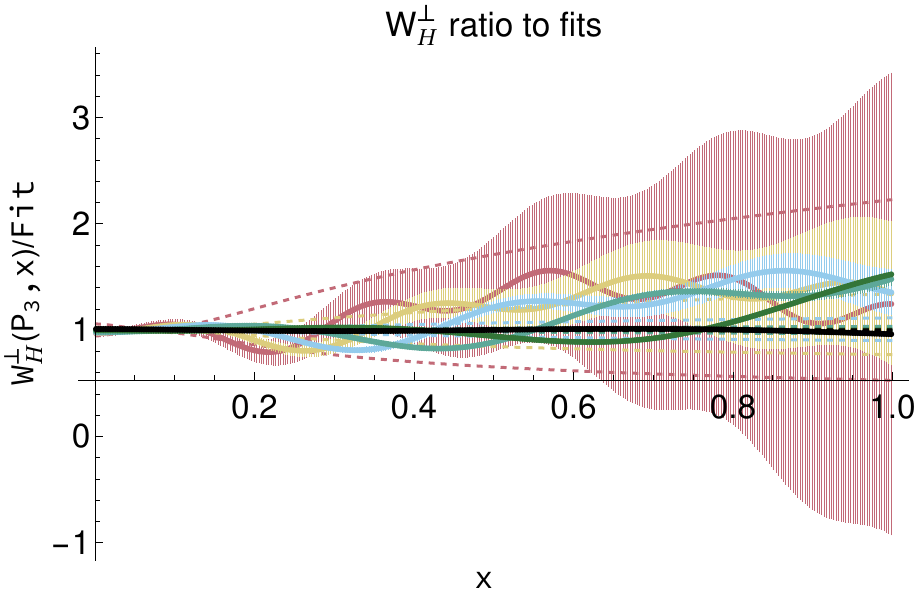}\includegraphics[width=0.5\textwidth]{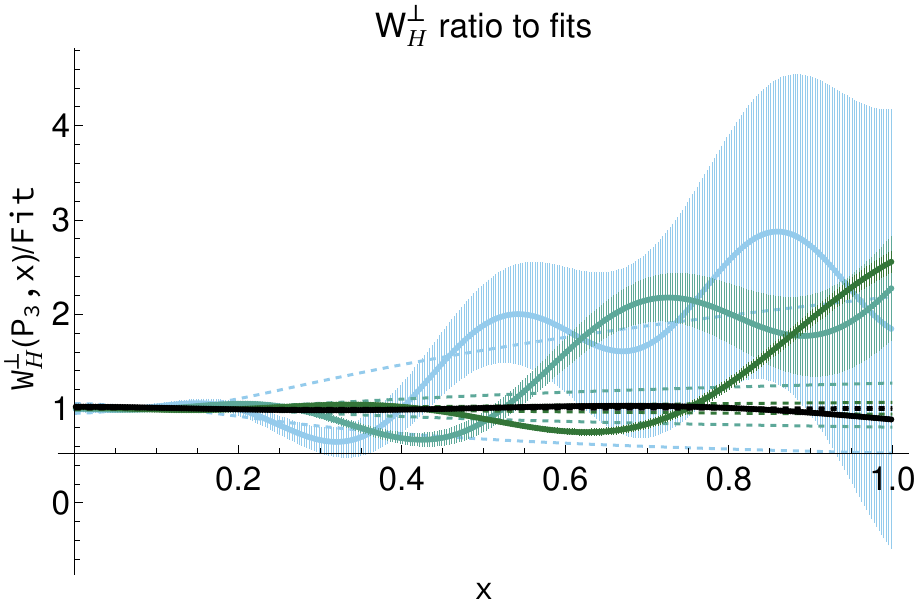}\\
  \includegraphics[width=\textwidth]{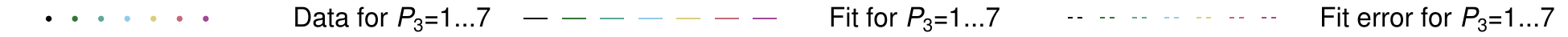}
 \caption{Fits and ratio to the fits for the quasi-PDFs $H_{W_2}$ (two top rows) and $W^\perp_H$ (two bottom rows), for the range of momenta where a fit was statistically possible. The top row and the third row show a direct comparison to the fits and the second and last row a ratio to the fit. The left column is set H and the right column set L.}
 \label{fits}
\end{figure}

A detailed example of the fitting procedure is shown in figure~\ref{fits}. There are two main obstacles for a fit with the simple functions~\prefr{fit1}{fit4}. One is the aforementioned finite volume effect in form of a constant offset in quasi-PDFs obtained from the scalar current~\pref{f0}. The remnant of this constant leads to an oscillatory behavior in $x$, which increases with momenta. By fitting the whole range of $x$ with non-oscillating fit ans\"atze, this averages out the oscillations at the expense of an enlarged error band. Thus, this covers the consequences of this lattice artifact, which would eventually vanish in the infinite volume limit.

The second is that quasi-PDFs are defined for all\footnote{We also used the data to check pseudo-PDFs~\cite{Cichy:2018mum}, which are naturally restricted to the physical $x$-range. However, there the propagation of statistical uncertainties yielded no useful signal.} $x$~\cite{Cichy:2018mum}. As a consequence, they will usually not vanish at $x=1$, even if the true PDF does. Indeed, we observe that with increasing $P_3$ some quasi-PDFs tend to vanish at $x=1$, and all become more suppressed towards $x=1$. In fact, due to the increase of statistical errors with increasing $P_3$, eventually all PDFs become consistent with zero at $x=1$ within errors. This feature is, to some extent, built into the fit forms~\prefr{fit2}{fit3}, which at finite $P_3$ force the fit away from the data.

The preferred fit forms of the different sources~\prefr{f1}{f2} and~\prefr{f4}{f5} indicate a qualitative difference between them. Correspondingly, we will provide both versions in the result section~\ref{s:pdf}, for the reader to gauge the associated systematic error.

Finally, we note that, up to lattice artifacts, there is good qualitative agreement between all three sets and quantitatively small differences between the parametrically matched pure gauge set L and the set F featuring dynamical fermions.

\section{PDFs}\label{s:pdf}

In this section, we provide our best estimates of the PDFs investigated. This should provide an overall idea of their shape and size, and thus a basis for the use in phenomenological calculations.

\begin{figure}
 \includegraphics[width=0.5\textwidth]{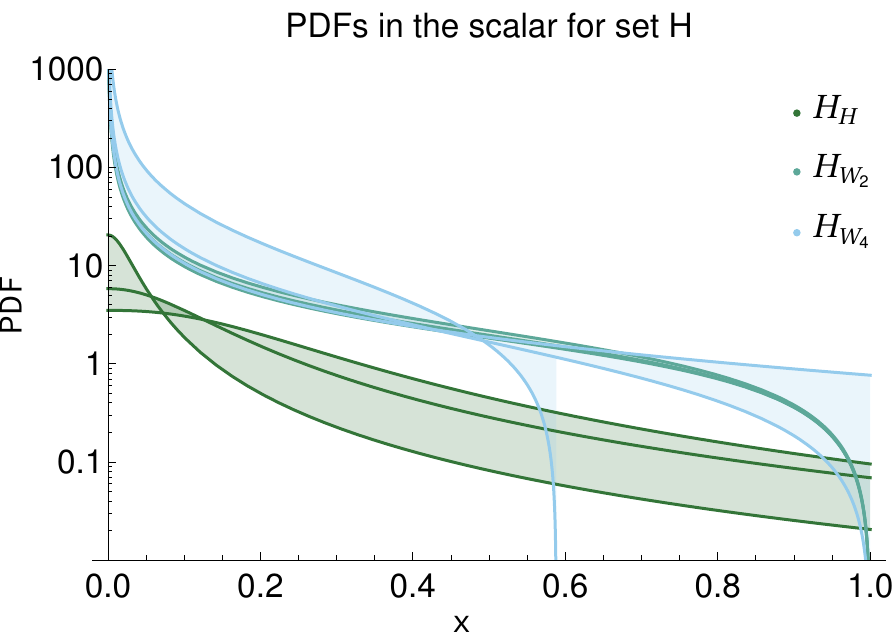}\includegraphics[width=0.5\textwidth]{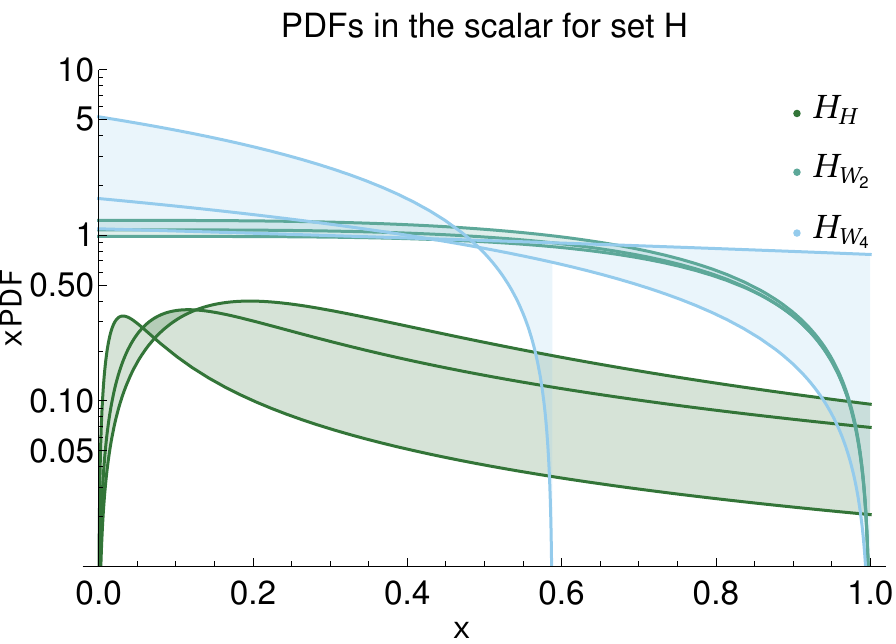}\\
 \includegraphics[width=0.5\textwidth]{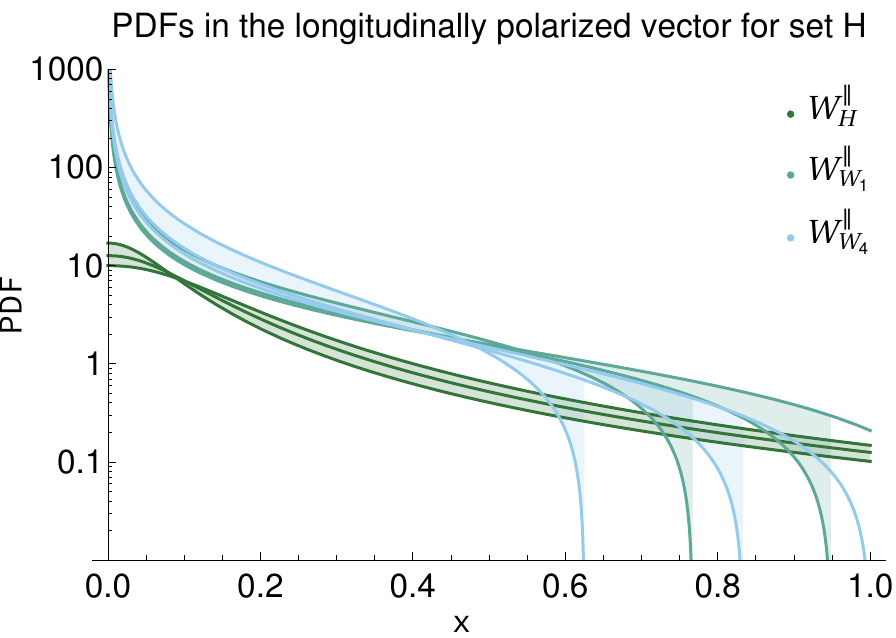}\includegraphics[width=0.5\textwidth]{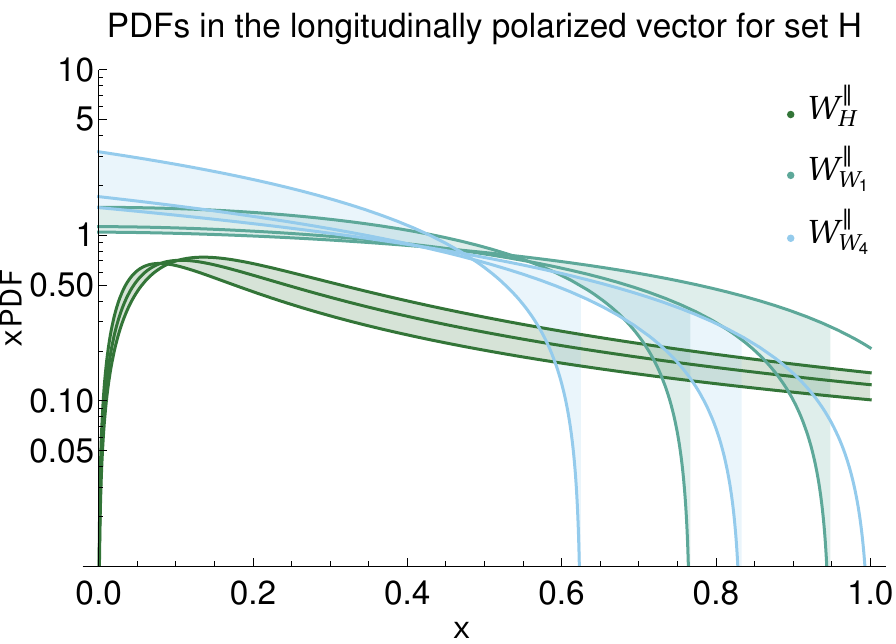}\\
 \includegraphics[width=0.5\textwidth]{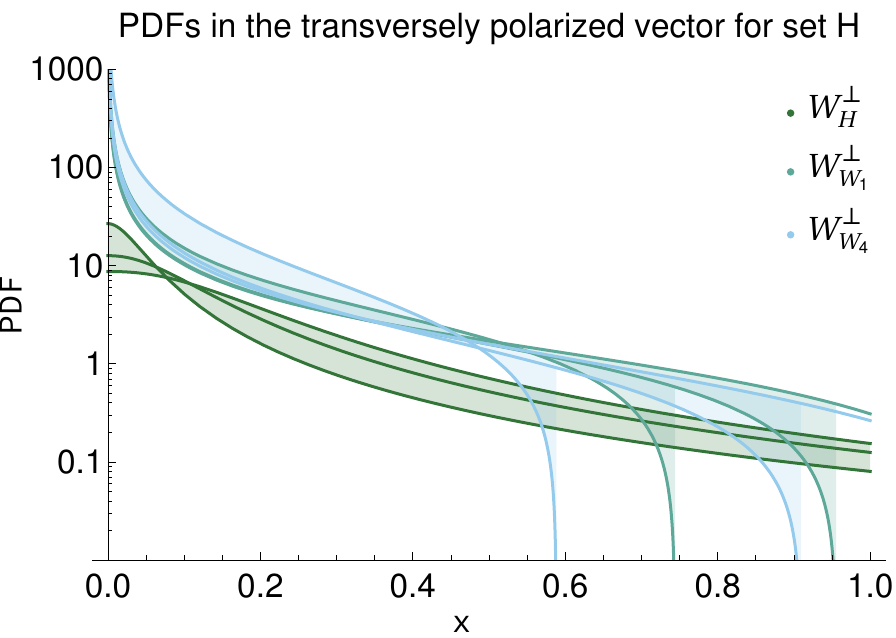}\includegraphics[width=0.5\textwidth]{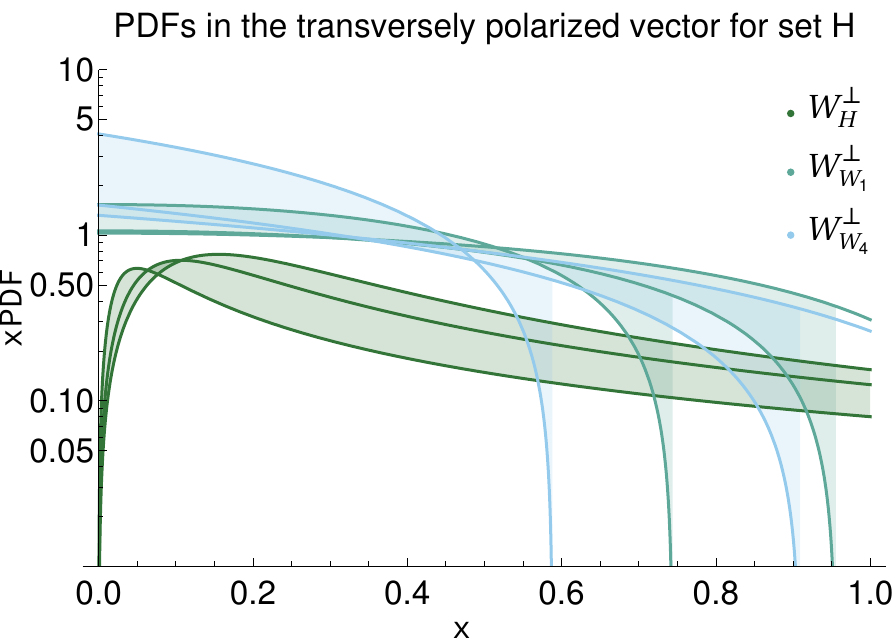}
 \caption{Final estimates for the PDFs for set H. The left column shows the PDFs themselves, the right column shows the PDFs multiplied by $x$. The top line are the PDFs for the scalar boson, the middle line for the longitudinally polarized vector boson, and the bottom for the transversely polarized vector boson.}
 \label{hpdf}
\end{figure}

\begin{figure}
 \includegraphics[width=0.5\textwidth]{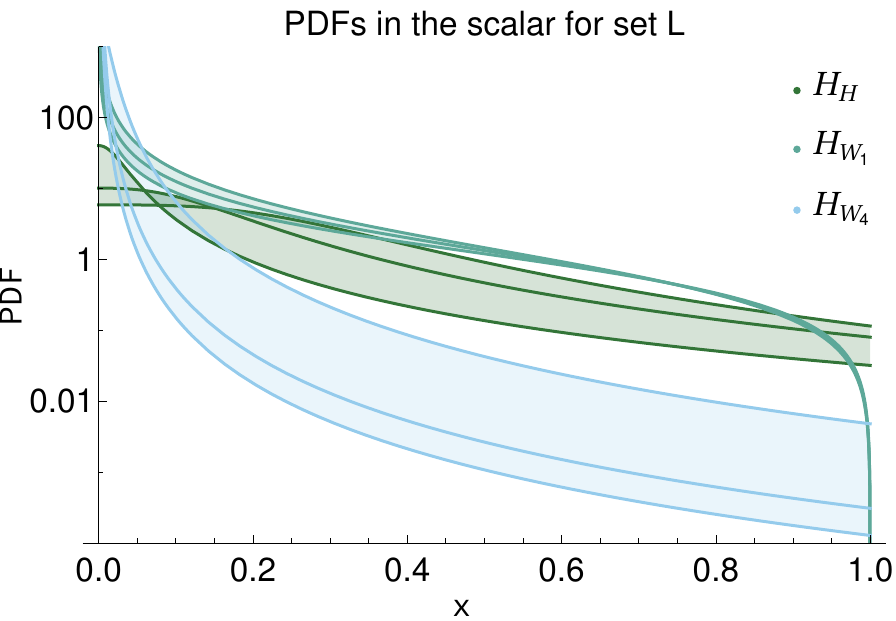}\includegraphics[width=0.5\textwidth]{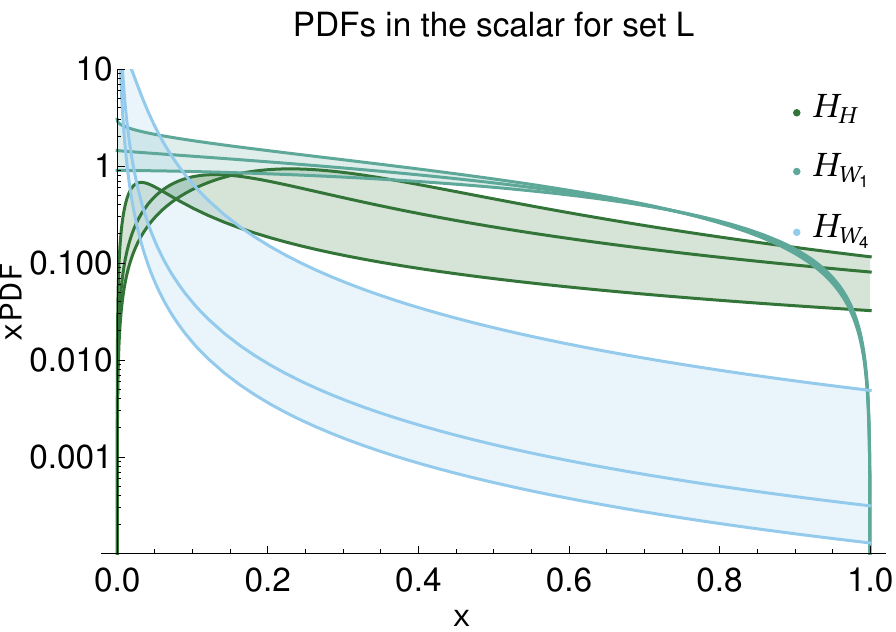}\\
 \includegraphics[width=0.5\textwidth]{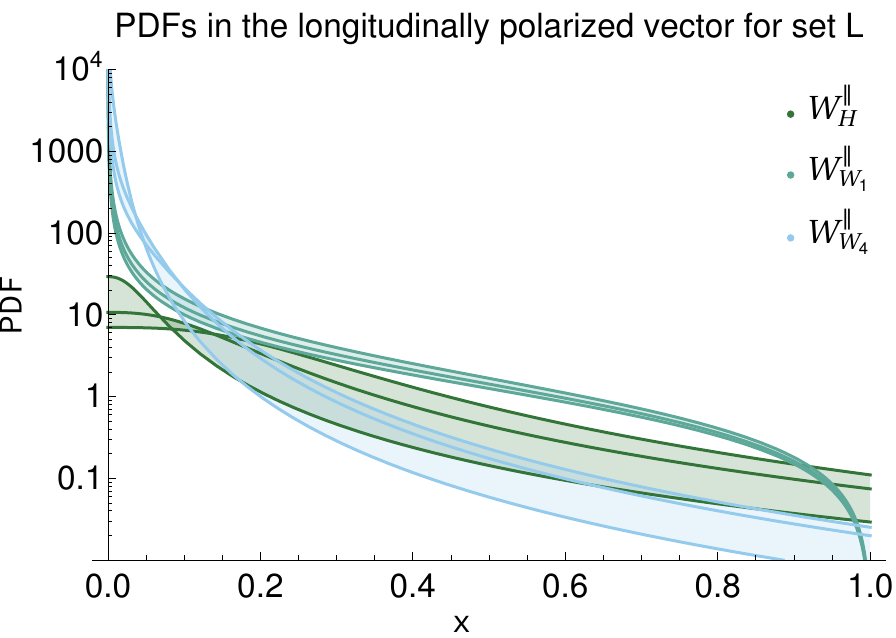}\includegraphics[width=0.5\textwidth]{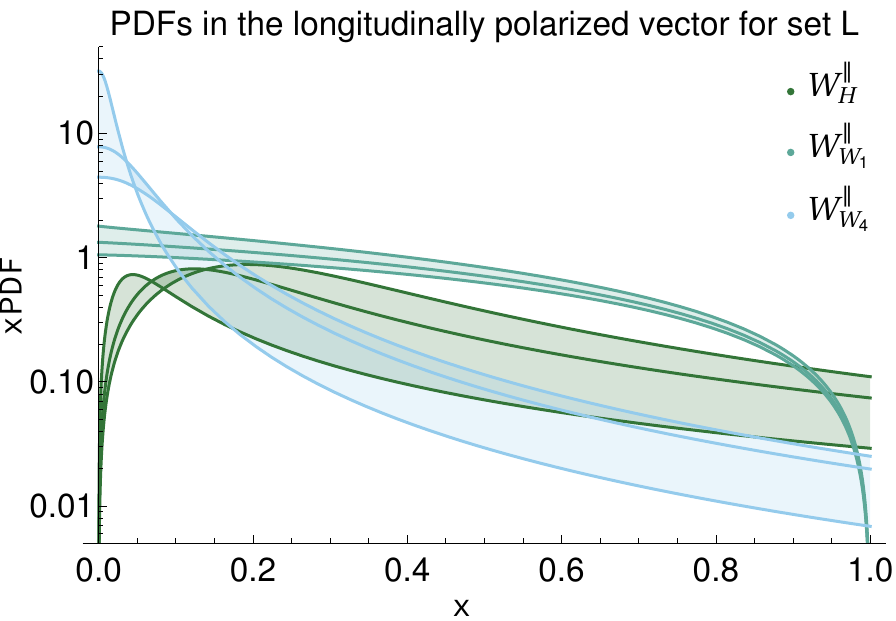}\\
 \includegraphics[width=0.5\textwidth]{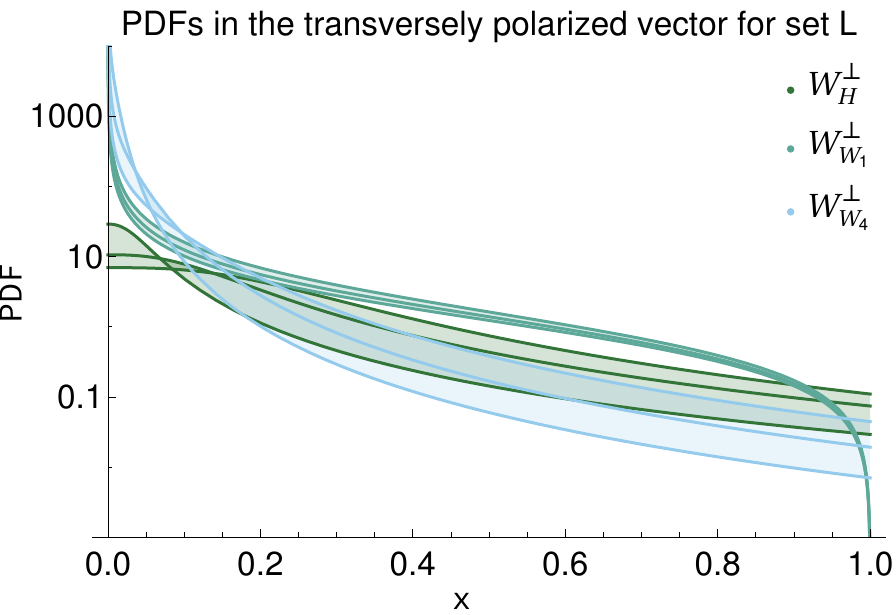}\includegraphics[width=0.5\textwidth]{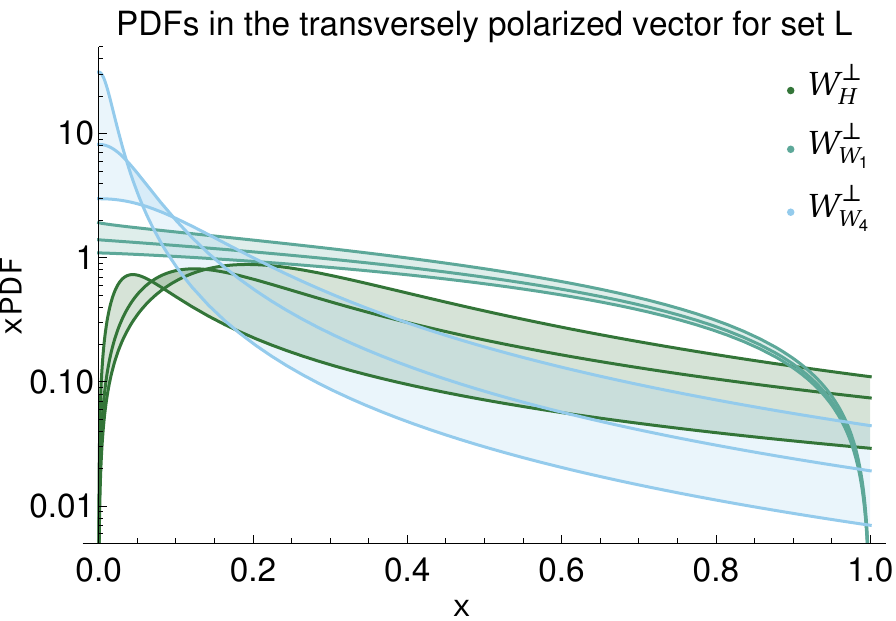}
 \caption{Final estimates for the PDFs for set L. The left column shows the PDFs themselves, the right column shows the PDFs multiplied by $x$. The top line are the PDFs for the scalar boson, the middle line for the longitudinally polarized vector boson, and the bottom for the transversely polarized vector boson.}
 \label{lpdf}
\end{figure}

\begin{figure}
 \includegraphics[width=0.5\textwidth]{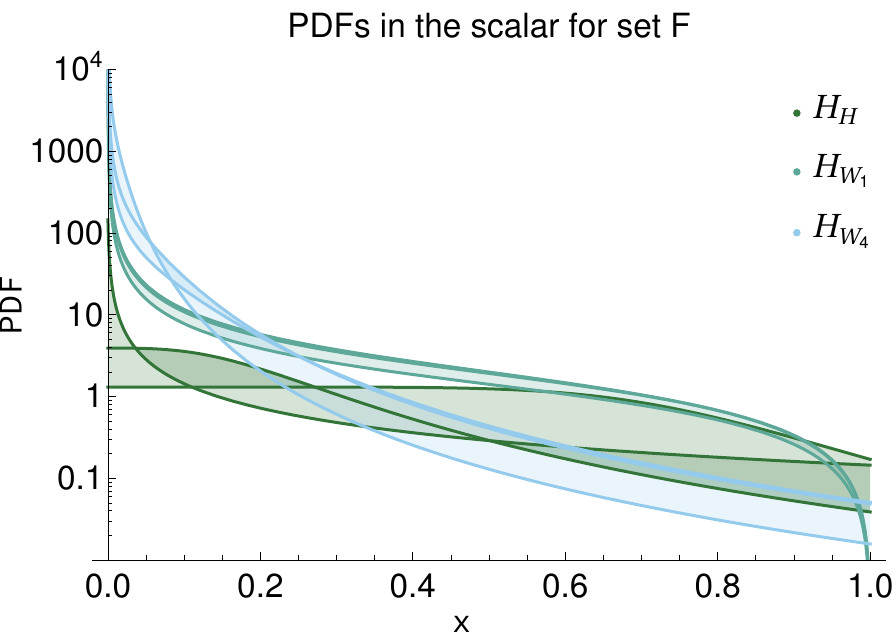}\includegraphics[width=0.5\textwidth]{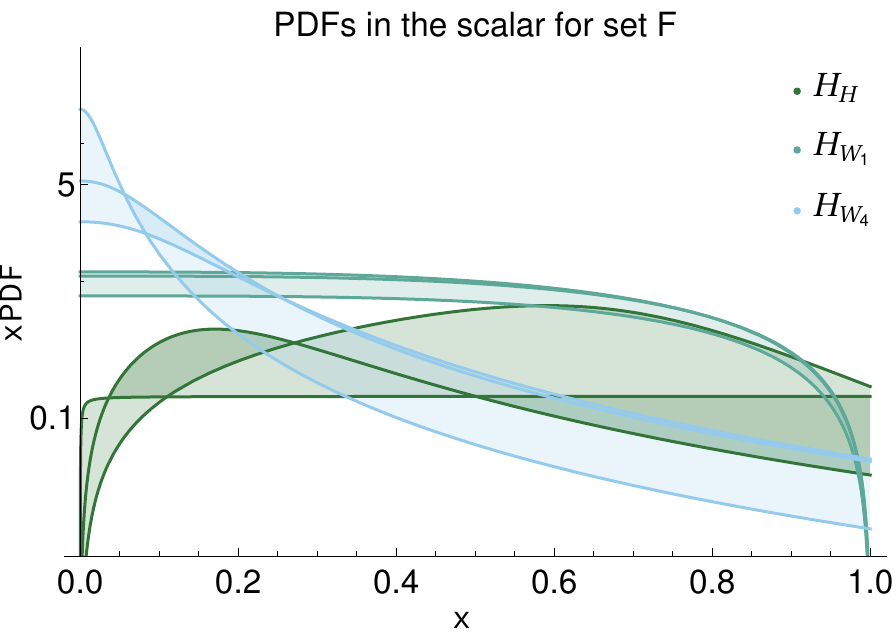}\\
 \includegraphics[width=0.5\textwidth]{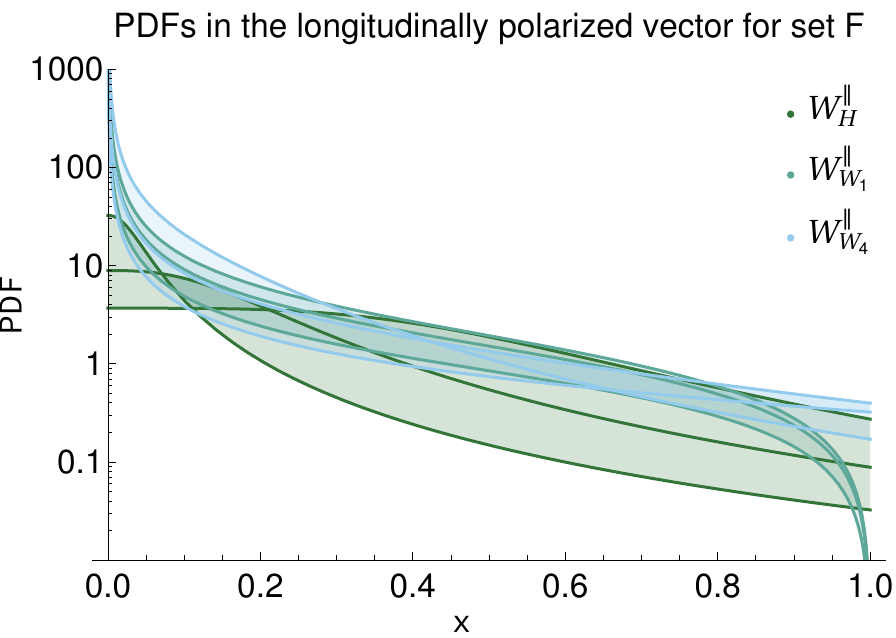}\includegraphics[width=0.5\textwidth]{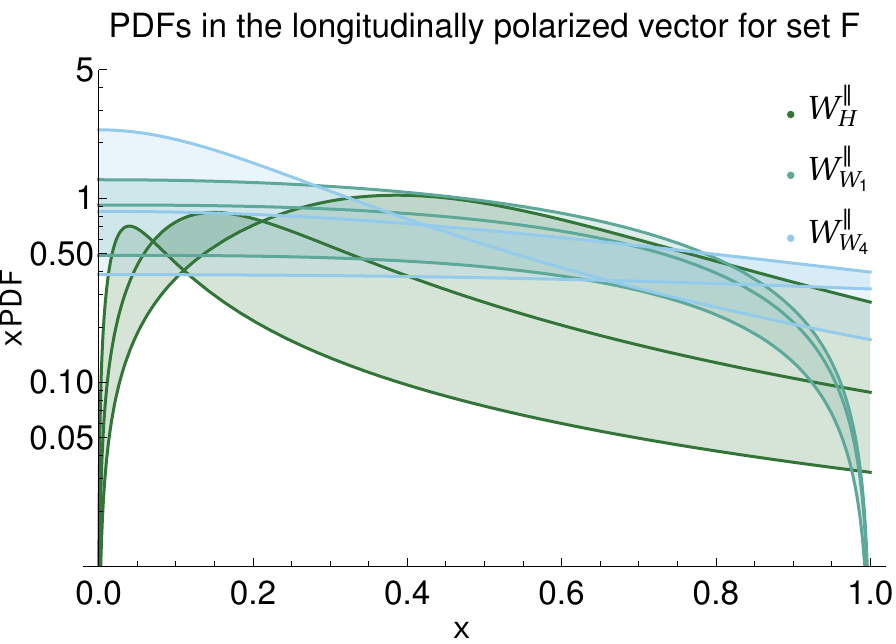}\\
 \includegraphics[width=0.5\textwidth]{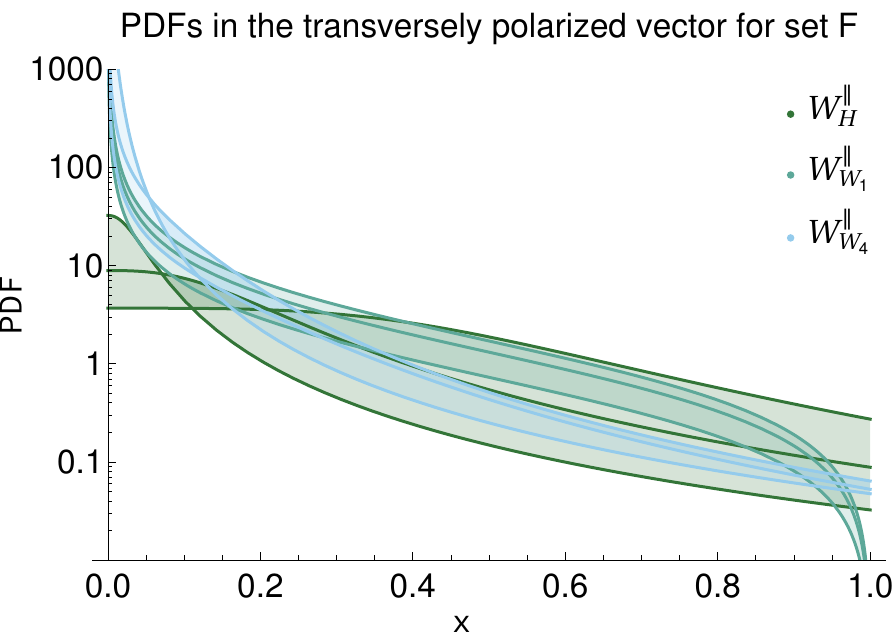}\includegraphics[width=0.5\textwidth]{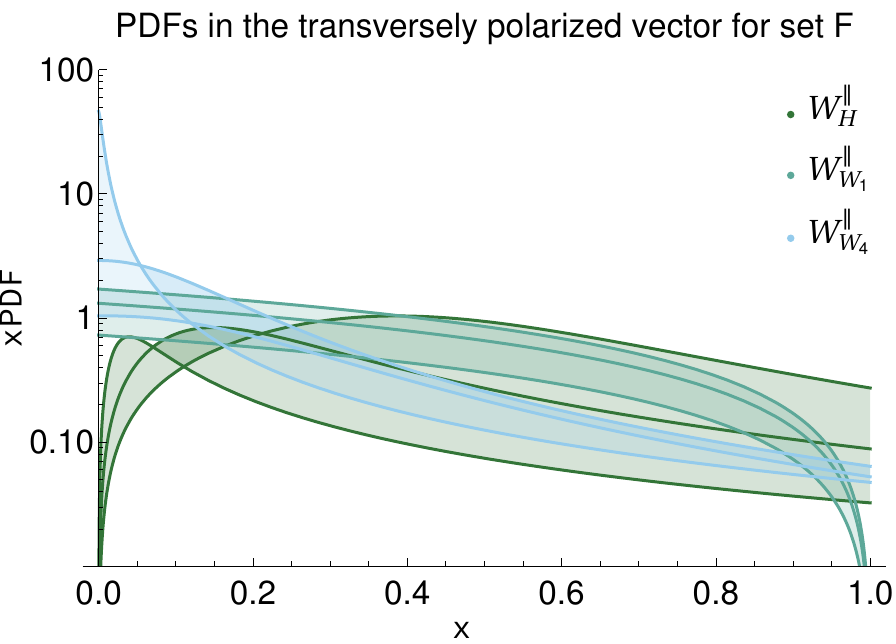}
 \caption{Final estimates for the PDFs for set F. The left column shows the PDFs themselves, the right column shows the PDFs multiplied by $x$. The top line are the PDFs for the scalar boson, the middle line for the longitudinally polarized vector boson, and the bottom for the transversely polarized vector boson.}
 \label{fpdf}
\end{figure}

These final estimates for the PDFs are shown in figures~\ref{hpdf},~\ref{lpdf}, and~\ref{fpdf} for the sets H, L, and F, respectively. The overall normalization has been fixed by the sum rules of section~\ref{s:normalization}.

We find that the qualitative features of the results do not depend on the set and that the Higgs PDF is never enhanced at small $x$, in strong contrast to quark PDFs. A similar saturation was already anticipated in phenomenological fits~\cite{Fernbach:2020tpa,Egger:2017tkd}. From a gauge-fixed perspective, this could be interpreted as that low-energy Higgs particles are absorbed in the condensate, and cannot be excited to participate in the hard interaction. Low-momentum modes will then rather be part of the condensate than dynamically available, even at very high energies. The gauge boson PDFs, on the other hand, are strikingly different and behave like $1/x$. In that they are closer to QCD-like PDFs. This also suggests that the suppression of the Higgs PDF is not (only) a mass effect, as the Higgs and the gauge bosons have a similar mass. 

Further, the PDFs obtained from the different currents for the gauge boson, the currents~\pref{f1} and~\pref{f4} in figures~\ref{hpdf} to~\ref{fpdf}, overlap within their error bands in almost all cases, with the results from $F_1$ usually with smaller errors. It should also be noted that the fit forms are different depending on the channel and current. However, we found that the fit form~\pref{fit2} pairs usually with the current~\pref{f1} and fit form~\pref{fit4} with the current~\pref{f4}. This is consistent with the anticipated universality~\cite{Zhang:2018diq}. Residual differences are most strongly pronounced in set L. 

While the question of normalization is subtle, as described in section~\ref{s:normalization}, we observe a trend that the Higgs PDF obtained from the stronger coupled set H is more heavily suppressed at larger $x$ compared to the sets L and F with weaker couplings and the W boson PDF on all other sets.

We further observe that the difference between the transversely polarized vector boson and longitudinally polarized vector boson are minor at best and that the difference between the scalar and the vector boson are also only moderate at large $x\gtrsim0.1$. The presence of dynamical, heavy fermions does not have a qualitative impact.

Probably the most interesting observations are that the all PDFs are modified substantially from their low-energy form~\prefr{tlpdf1}{tlpdf4} (or including~\pref{tlpdf5} instead of~\pref{tlpdf4}), and that the Higgs PDF behaves qualitatively different from the gauge boson PDFs. In particular, at low $x$, the gauge bosons dominate massively. In this context, it is necessary to keep in mind that our approximations will only reasonably work at large energies compared to the masses. Thus, the latter statement is likely only accurate at energy scales $xE\gg m_H$. Already for a modest value of $x=0.1$ this requires multi-TeV interactions, the region where non-linear effects of the electroweak interactions are relevant~\cite{deBlas:2025gyz,Maas:2022gdb}.

\section{Conclusion}\label{s:conclusion}

We have provided a first lattice investigation of weak PDFs in a reduced Standard Model, which contains only the Higgs and the $W$/$Z$, and a single exploratory look at the consequences of including heavy, vector-like fermions. This constitutes the first ever non-perturbative lattice investigation of PDFs for this theory, and thus the first ever non-perturbative investigation of the weak PDFs considered to be highly relevant for future experiments~\cite{deBlas:2025gyz,Maas:2022gdb,Egger:2017tkd,Bauer:2017isx,MoralFigueroa:2025cev}.

Our results suggest that the PDFs are relatively similar between the physical scalar and vector particles of the theory for $x>0.1$. However, the Higgs PDF is substantially suppressed at low $x$, compared to the gauge boson PDF.

We have thoroughly investigated systematic effects. In particular, we have found that we have good control over the finite volume effects and the extrapolation of $P_3\to\infty$, the Bjorken limit. We examined the impact of different currents for the matrix elements, and studied thoroughly multiple fit ansatz choices, and found no qualitative, and only small quantitative, effects. While we did not yet investigate systematically discretization effects, there are strong indications for them to be small~\cite{Maas:2014pba}. Also, our two ensembles H and L, while having slightly different Higgs masses, do have substantially different lattice spacing, but this did not lead to any qualitative, and only mild quantitative, differences. Of the remaining uncertainties, the largest effects stem from the choice of current, and the infinite-momentum extrapolation. Moreover, we neglected effects due to factorization, and thus our results will likely only be leading-order accurate.

There is still a long way to go before these results reach a similar precision as those employed in QCD calculations. Introducing factorization and improving statistical uncertainties are likely the most relevant candidates for future research. But for now, with experiments only starting to probe the relevant domain,  our results can certainly serve as a first input to exploratory calculations based on the physical asymptotic states.

An interesting possibility will also be to evolve these results towards lower $Q$, to pinpoint the necessary valence contribution, which is usually assumed to be zero~\cite{Bauer:2017isx} or trivial~\cite{Egger:2017tkd,Maas:2022gdb}. That this will be a significant effect becomes evident by considering results for vector boson scattering~\cite{Jenny:2022atm}, which at low energies cannot be captured by the trivial PDFs of section~\ref{s:normalization}. Conversely, it will be also very interesting to use the present results for predictions of high-energy inclusive inelastic cross sections, e.\ g.\ again in vector boson scattering. There, deviations are also expected~\cite{Maas:2022gdb}, and their determination using lattice methods is within reach~\cite{Martins:2026eni}.

\section*{Acknowledgements}
We are grateful to S.~Zafeiropoulos for helpful discussions. This research was funded in whole or in part by the Austrian Science Fund (FWF) [\href{https://www.fwf.ac.at/en/research-radar/10.55776/PAT6443923}{10.55776/PAT6443923}]. The authors acknowledge support from the European Union~(EU) COMETA COST Action~(\href{https://www.cost.eu/actions/CA22130}{CA22130}).  The computational results have been achieved using the Austrian Scientific Computing (ASC5) infrastructure and the Graz Scientific Cluster (GSC1). All matrix elements and quasi-PDFs can be downloaded from~\cite{data_release}.





\bibliography{bib.bib}

\end{document}